\documentclass[intlimits,twoside,a4paper]{article}
\usepackage{amsmath,amssymb}
\usepackage{graphicx}
\usepackage{wrapfig}
\usepackage[T2A]{fontenc}
\usepackage[cp1251]{inputenc}
\usepackage{color}

\usepackage{cmpj2}

\issue{2013}{16}{1}{13004}

\doinumber{10.5488/CMP.16.13004}

\title[Fokker-Planck equation with memory]{Fokker-Planck equation with memory: the crossover from ballistic to diffusive processes in many-particle systems and incompressible media}

\author[V. Ilyin, I. Procaccia, A. Zagorodny]{V. Ilyin\refaddr{ad1}, I. Procaccia\refaddr{ad1}, A. Zagorodny\refaddr{ad2}}

\addresses{\addr{ad1}Department of Chemical Physics, The Weizmann
Institute of Science, 76100 Rehovot, Israel%
\addr{ad2}  Bogolyubov Institute
for Theoretical Physics of the National Academy of Sciences of Ukraine, \\ 03680 Kyiv, Ukraine}

\authorcopyright{V. Ilyin, I. Procaccia, A. Zagorodny, 2013}
\date{Received  July 24, 2012}  

\newcommand{\ud}{\mathrm{d}}
\newcommand{\wt}[1]{\widetilde{#1}}

\begin{document}

\maketitle

\begin{abstract}

The unified description of diffusion processes that cross over from a
ballistic behavior at short times to normal or anomalous diffusion (sub- or
superdiffusion) at longer times is
constructed on the basis of a non-Markovian generalization of the Fokker-Planck
equation. The necessary non-Markovian kinetic coefficients are determined by the
observable quantities (mean- and mean square displacements).
Solutions of the non-Markovian equation describing
diffusive processes in the physical space are obtained. For long
times, these solutions agree with the predictions of the continuous
random walk theory; they are, however, much superior at shorter times
when the effect of the ballistic behavior is crucial.

\pacs 05.40 Fb, 05.40 Jc, 51.10 +y
\end{abstract}

\section{ Introduction}

Although particle diffusion processes have been studied for
about two centuries~\cite{P05}, there are still some subtle issues that need to be faced
at the present time. In this paper, we point out these subtle difficulties and offer ways
to overcome them.

The quantitative theory of diffusion processes begins in 1855 with the phenomenological solution of the diffusion problem by Fick. Fick employed an empirical definition of the particle flux
through the surface of a subvolume (Fick's first law) and the continuity
equation which reflects the conservation of particles.
This combination results in the diffusion equation (the Fick's second law)
which defines the time evolution of the probability distribution
function (PDF) of particle concentration. The variance of this PDF grows in time according  to
\begin{equation}
\langle r^2\rangle_t\sim Dt,
\label{var1}
\end{equation}
where $D$ is the diffusion coefficient which in general depends on the particles and medium in which they
diffuse. The equation solved by the PDF is the classical diffusion equation
\begin{equation}
\frac{\partial f(r,t)}{\partial t}
=D\frac{\partial^2 f(r,t)}{\partial r^2} \, .
\label{cldiff}
\end{equation}
The solution of this equation with the initial condition
\begin{equation}
f(r,t=0)=\delta(r),
\label{incond}
\end{equation}
where $\delta(x)$ is the Dirac delta-function, reads
\begin{equation}
f(r,t)=\frac{\exp\left[-r^2/(4Dt)\right]}{\sqrt{\pi D t}}\,.
\label{classol}
\end{equation}
This function is normalized to unity $\int_0^{\infty}f(r,t) \ud r=1$ and
its variance agrees with equation~(\ref{var1}).

The first subtle difficulty that one needs to pay attention to is that this solution
allows particles to arrive at arbitrary distances from the origin in finite, and even infinitesimal time.
Moreover, in many cases, the processes that exhibit the law (\ref{var1}) for times larger than some time $t_\mathrm{c}$, possess a different behavior at short times, i.e.,
\begin{equation}
\langle r^2\rangle_t\sim D_2t^2, \qquad \text{for} \qquad t \ll t_\mathrm{c}\, .
\label{varbal}
\end{equation}
This behavior is referred to as ``ballistic''.
Recently, due to the development of the measuring equipment, it has become
possible to observe the ballistic motion and the transition to normal diffusion in
accordance with equation~(\ref{var1})~\cite{LJTKFF05,HCTLJRF11,P11}. Therefore,
the correct theory for the PDF should satisfy the two asymptotic limits given by
equation~(\ref{varbal}) and, in general, by equation~(\ref{var1}) simultaneously.

The cure for both the infinite speed problem and for the ballistic regime in 1 dimension was proposed by Davydov~\cite{D34}, who introduced an explicit time interval $t_\mathrm{c}$ of the
mean free path. In his approach, the diffusion problem is reduced to
a solution of the well known telegraph equation (see, e.g.,~\cite{MF53}).
In contrast to the parabolic nature of Fick's second law,  the telegraph
equation is of a hyperbolic type, resulting in a
propagation of the front of the PDF with finite velocity. However, we reiterate  that it was
stressed in~\cite{D34} that the telegraph equation is correct only in the
one-dimensional case. The equivalent treatment for 2 and 3 dimensional diffusion does not exist in the
literature. One of the main results of this paper (cf. section~\ref{hyd}) is a way to achieve the same
in dimensions higher than 1.

Another issue that needs to be discussed is that the classical diffusion law equation~(\ref{var1})
is not universally obeyed in Nature. Since the classical
work by Hurst~\cite{51Hur} on the stochastic discharge of reservoirs and
rivers, Nature has offered us a large number of examples of diffusion
processes which are `anomalous' in the sense that an observable $X$
diffuses in time, so that its variance time dependence is
\begin{equation}
\langle \Delta X^2\rangle_t\sim D_\alpha  t^\alpha \ , \qquad t\gg t_\mathrm{c}\,,
\label{var2}
\end{equation}
where $\alpha\ne 1$. We expect $\alpha\ne 1$
when the diffusion steps are correlated, with persistence for $\alpha>1$ and
anti-persistence for $\alpha<1$~\cite{MN68}.
This law is usually valid only at long times and for $t\ll t_\mathrm{c}$ we  may again
find ballistic behavior. All the issues discussed above for classical diffusion reappear
in the context of anomalous diffusion, both in 1 and in higher dimensions. The bulk of this paper
will deal with establishing the methods to achieve a consistent theory for the PDF that is valid
for all times.

As a preparation for more complex situations,  in section~\ref{telegraph} we review the telegraph
equation that regularizes all the issues raised above in 1 dimension. In section~\ref{Fokker} we
generalize the methodology of the telegraph equation to Fokker-Planck equations with
arbitrary non-local kernels in time. In the same section we show how to relate this kernel
to the observable mean square displacement of a diffusive process. In the next section
we turn to the Langevin equation for a guidance how to compute the mean square displacement to
uniquely  determine the associated Fokker-Planck equation. In the next Section we generalize the methods
used in this paper to the 3-dimensional case taking into account hydrodynamic interactions. The last section offers a summary and conclusions.

\section{Review: the telegraph equation in 1 dimension.}
\label{telegraph}
In order to develop  a diffusion model with possible finite speed of propagation,
it is necessary to simultaneously take into account   the mean particle
velocity $c$ and the time length $t_\mathrm{c}$ of the mean free path~\cite{D34}.
Then, the diffusion process is defined by the following equation
\begin{equation}
t_\mathrm{c}\frac{\partial^2 f(r,t)}{\partial t^2}+\frac{\partial f(r,t)}{\partial t}
=D\frac{\partial^2 f(r,t)}{\partial r^2}\,,
\label{tlg}
\end{equation}
where the diffusion coefficient $D=c^2t_\mathrm{c}$.

In the limit $t_\mathrm{c}\to 0$, the time evolution of the PDF is possible only if
$D\ne 0$, e.g., the mean particle velocity $c\to \infty$. In this case,
equation~(\ref{tlg}) is reduced to the classical diffusion equation (\ref{cldiff})
and the solution of this equation with the initial condition (\ref{incond}) is
equation~(\ref{classol}). This solution is normalized to unity and its variance agrees with
equation~(\ref{var1}).

The delta-function initial condition can be considered as a limit of a sequence of
continues functions, for example
\begin{eqnarray}
\delta(x)&=&\lim_{\alpha\to \infty} \delta(x,\alpha),\nonumber \\
\delta(x,\alpha)&=&\frac{\alpha}{\sqrt{\pi}}\re^{-\alpha^2x^2}.
\label{delt}
\end{eqnarray}
Comparing with equation~(\ref{cldiff}) we conclude that normal diffusion is an inverted
process to the limit defined by equation~(\ref{delt}), the initial delta-function
spreads in space when time increases, keeping the center of mass at rest at the origin.

In the limit $t_\mathrm{c}\to \infty$, equation~(\ref{tlg}) transforms to the wave equation
\begin{equation}
\frac{\partial^2 f(r,t)}{\partial t^2}=
c^2\frac{\partial^2 f(r,t)}{\partial r^2}\,.
\label{wave}
\end{equation}
Its solution with the initial condition defined by equation~(\ref{incond}) reads
\begin{equation}
f(r,t)=\delta(ct-r) \ .
\label{solvwave}
\end{equation}
This solution looks as though the initial conditions propagate with the
velocity $c$ without changing the shape of the distribution.

\begin{figure}[!b]
\centering
\includegraphics[width=0.38\textwidth]{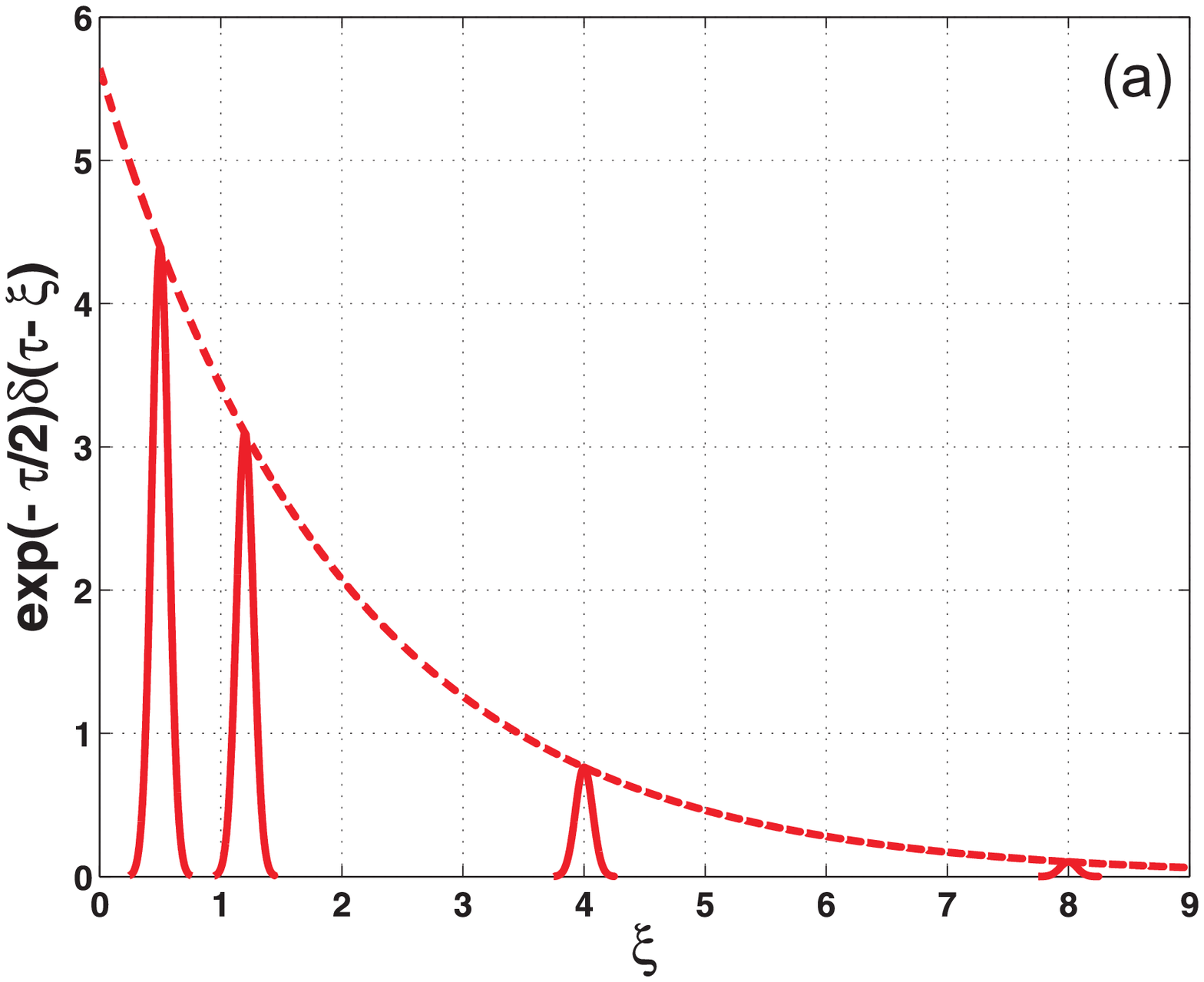}
\hspace{1cm}
\includegraphics[width=0.4\textwidth]{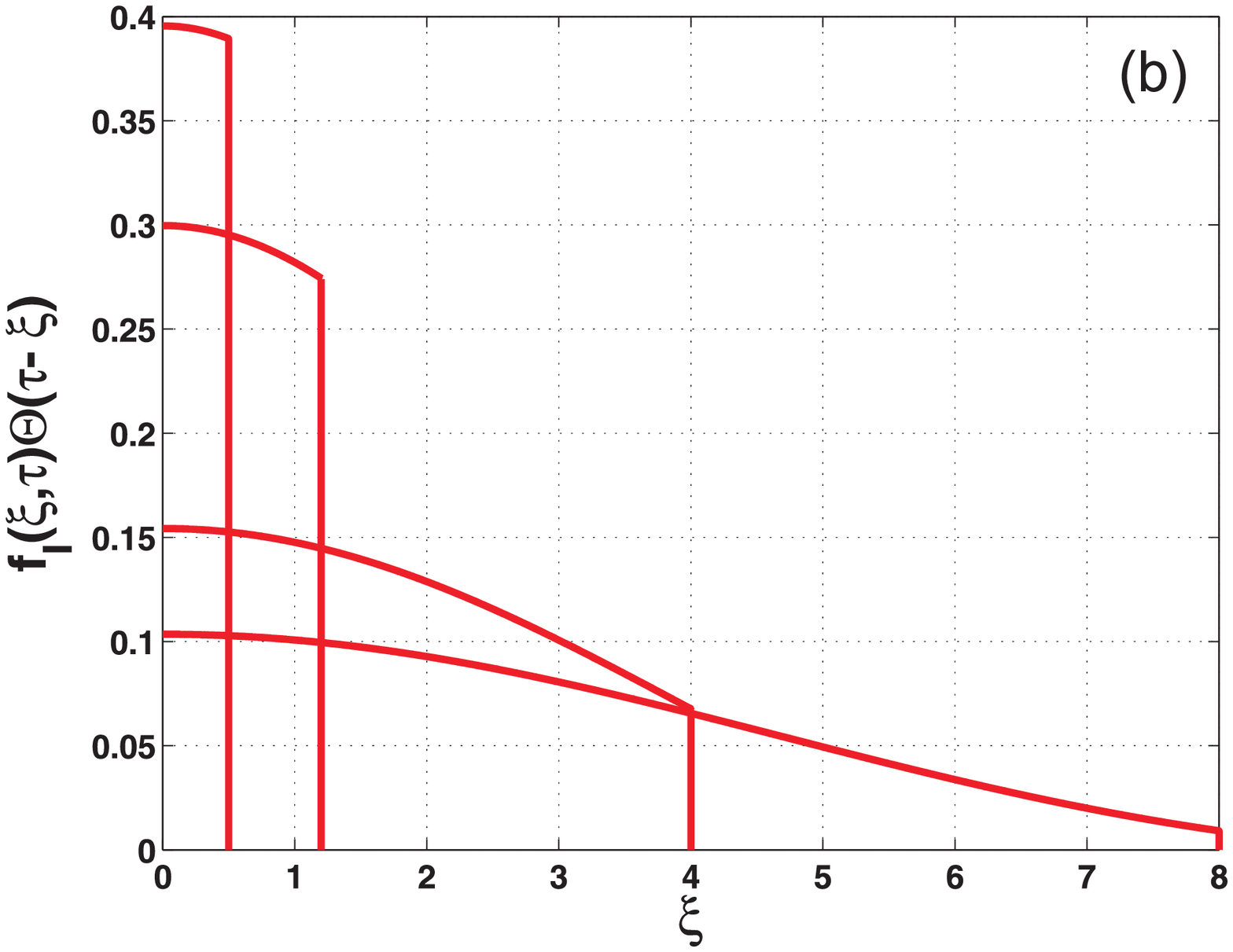}\\
\includegraphics[width=0.4\textwidth]{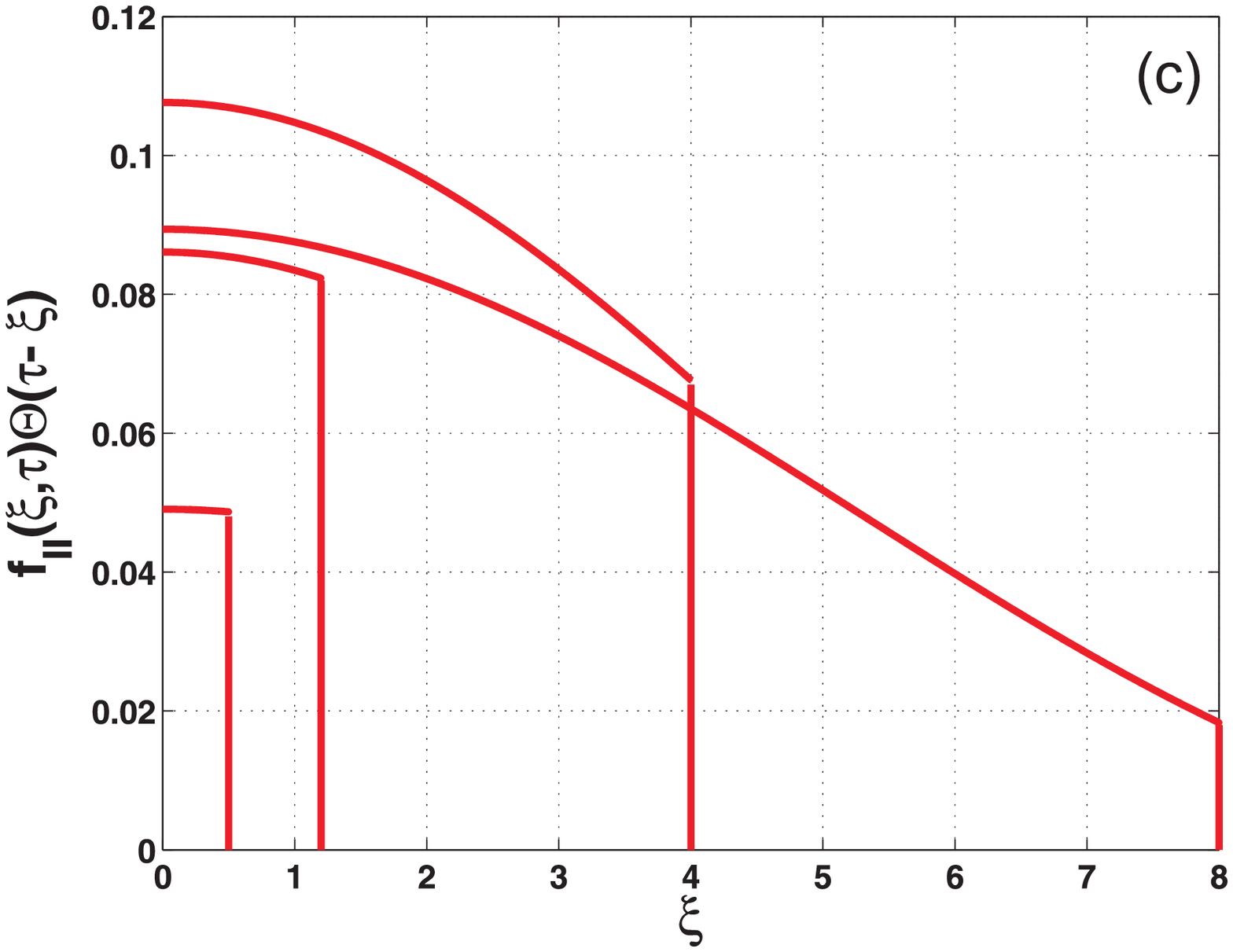}
\caption{(Color online) The three contributions to the solution of the telegraph equation PDF [equation~(\ref{soltlg})]
for different time intervals $\tau=$ 0.5, 1, 2, 4, 8. Propagating
contribution is shown in panel (a). The delta-function is graphically
represented by narrow Gaussians [see equation~(\ref{delt})]. The functions
$f_{I,II}(\xi,\tau)\Theta(\tau-\xi)$ [see equations~(\ref{f1}) and (\ref{f2})]
are shown in panels (b) and (c).}
\label{fig1}
\end{figure}

For an arbitrary value of $t_\mathrm{c}$, the solution of
the telegraph equation is given by three contributions
(see, e.g.,~\cite{IZ10}  and the references
cited therein)
\begin{eqnarray}
f(\xi,\tau)&=&\re^{-{\tau}/{2}}\delta(\tau-\xi) 
+\left[f_{I}(\xi,\tau)+f_{II}(\xi,\tau)\right]
\Theta(\tau-\xi) \, .
\label{soltlg}
\end{eqnarray}
Here, the two dimensionless variables are the normalized time $\tau=t/t_\mathrm{c}$ and the
normalized length $\xi=r/\sqrt{Dt_\mathrm{c}}$.  $\Theta(x)$ is the Heaviside function.
The variance of this distribution reads
\begin{equation}
\langle \xi^2\rangle_t=2\left[\tau -(1-\re^{-\tau})\right].
\label{vartlg}
\end{equation}
This form clearly crosses over from ballistic to normal diffusion as $\tau$ increases.
The interpretation of the three contributions is as follows: the first term corresponds to the damped ballistic propagation of the kind defined by equation~(\ref{solvwave}). The next two terms
exist only inside the compact region that expands with the velocity $c$. The functions
$f_{I}(\xi,\tau)$ and $f_{II}(\xi,\tau)$ are given by
\begin{equation}
f_{I}(\xi,\tau)=\frac{\exp(-\tau/2)}{2}
I_0\bigg(\frac{1}{2}\sqrt{\tau^2-\xi^2}\bigg)
\label{f1}
\end{equation}
and
\begin{equation}
f_{II}(\xi,\tau)=\tau\frac{\exp(-\tau/2)}{2}
\frac{ I_1\bigg(\frac{1}{2}\sqrt{\tau^2-\xi^2}\bigg)}{\sqrt{\tau^2-\xi^2}}\,,
\label{f2}
\end{equation}
where $I_\nu(z)$ are the modified Bessel functions of the first kind.
The asymptotic behavior of these functions is defined by
$I_{\nu}(z)_{z\to\infty}\sim \exp(z)/\sqrt{2\pi z}$, therefore, in the long
time limit $\tau\gg 1$, equation~(\ref{soltlg}) is reduced to the solution of
the classical diffusion equation given by equation~(\ref{classol}).

Various terms in the solution of the telegraph equation, equation~(\ref{soltlg}),
are shown in figure~\ref{fig1}. Physically, the first term describes the part of the initial mass of
diffusers at the origin that shoots out as a propagating solution which decreases exponentially in time, thus providing more and more mass to the
diffusive part of the transport process. The function
$f_{I}(\xi,\tau)\Theta(\tau-\xi)$ represents the spreading of diffusers whose mass peaks
at the origin without participating in the propagation of the first function.
Note that this solution is zero at $t=0$, but it is finite for any positive time. The third term  $f_{II}(\xi,\tau)\Theta(\tau-\xi)$ is seen to first increase in weight and in amplitude and later
on to decrease. It stems from the existence of a finitely rapid front that does not allow the unphysical
infinite speed of propagation that characterizes the solution (\ref{classol}). Of course, the
sum of the three contributions is normalized.

\section{From the telegraph equation to the time-nonlocal Fokker-Planck \\ equation}
\label{Fokker}
In order to motivate the use of the Fokker-Planck equation with time non-local kernels, we rederive
the telegraph equation in the following way:
let us begin with the continuity equation in the following form:
\begin{equation}
\frac{\partial f(r,t)}{\partial t}=-\frac{\partial {\bf \Gamma}(r,t)}
{\partial r} \, .
\label{cont}
\end{equation}
Here, ${\bf \Gamma}(r,t)$ is the particle flux. In its turn, this flux satisfies the equation~\cite{D34}
\begin{equation}
\frac{\partial {\bf \Gamma}(r,t)}{\partial t}=-\frac{{\bf \Gamma}(r,t)-
{\bf \Gamma}_0(r,t)}{t_\mathrm{c}}\,,
\label{Tflux}
\end{equation}
where ${\bf \Gamma}_0(r,t)$ is the same as  the first Fick's law
\begin{equation}
{\bf \Gamma}_0(r,t)=-D\frac{\partial f(r,t)}{\partial x}\,.
\label{Fik1}
\end{equation}
The solution of equation~(\ref{Tflux}) is given by~\cite{C48,APLA08}
\begin{equation}
{\bf \Gamma}=-\frac{D}{t_\mathrm{c}}\int\limits_{0}^{t}\exp\left(-\frac{t-t^\prime}
{t_\mathrm{c}}\right)\frac{\partial^2 f(r,t^\prime)}{\partial r^2}\ud t^\prime.
\label{dinflux}
\end{equation}
Substitution of this equation into equation~(\ref{cont}) yields the time
nonlocal diffusion integro-differential equation
\begin{equation}
\frac{\partial f(r,t)}{\partial t}=\frac{D}{t_\mathrm{c}}\int\limits_{0}^{t}
\exp\left(-\frac{t-t^\prime}
{t_\mathrm{c}}\right)\frac{\partial^2 f(r,t^\prime)}{\partial r^2}\ud t^\prime.
\label{difnl}
\end{equation}
After time differentiation, this equation is reduced to the telegraph
equation defined by equation~(\ref{tlg}).

Observe now~\cite{IZ10} that equation~(\ref{difnl}) is a particular case of the time-nonlocal Fokker-Planck
equation,
\begin{equation}
\frac{\partial f(r,t)}{\partial t} = \int_0^t  W(t-t^\prime)
\frac{\partial ^2 f(r,t^\prime)}{\partial r^2}\ud t^\prime
\ ,
\label{FPpdf}
\end{equation}
where $W(t)$ is the kernel responsible for the non-Fickian behavior of
the diffusion process. The Laplace transform of the solution of this equation
is given by~\cite{S02}
\begin{equation}
\wt{f}(r,s)=\frac{1}{\wt{W}(s)}\cdot \wt{P}\left(r,{s}/{\wt{W}(s)}\right),
\label{formsol}
\end{equation}
where the function $\wt{P}(r,s)$ is the Laplace transform of the
solution of the auxiliary   equation with the same  initial condition
\begin{equation}
\frac{\partial}{\partial t}P(r,t)=\frac{\partial^{2}}{\partial r^{2}}P(r,t)\,.
\label{Meq}
\end{equation}
which is identical with equation~(\ref{cldiff}). The Laplace transform of the
solution corresponding to the initial condition defined by equation~(\ref{incond})
follows from equation~(\ref{classol}) with $D=1$
\begin{equation}
\wt{P}(r,s)=\sqrt{\frac{2}{\pi}}\frac{rK_{1/2}\left(r\sqrt{s}\right)}{\left(r\sqrt{s}\right)^{1/2}},
\label{Plpl}
\end{equation}
where $K_{1/2}(z)$ is the modified Bessel function of the third kind.
Substitution of equation~(\ref{Plpl}) into equation~(\ref{formsol}) yields the solution of
equation~(\ref{FPpdf})
\begin{equation}
\wt{f}(r,s)=\sqrt{\frac{2}{\pi}}\frac{r}{\wt{W}(s)}
\frac{K_{1/2}\left(r\sqrt{s/\wt{W}(s)}\right)}{\left(r\sqrt{s/\wt{W}(s)}\right)^{1/2}}\,.
\label{NMsol}
\end{equation}

The Laplace transform of an even moment of the  distribution is defined by
\begin{equation}
\langle r^{2m}\rangle_s=\int\limits_0^\infty \wt{f}(x,s) x^{2m} \ud x,
\qquad  m=1,2, \ldots \,.
\label{evnmom}
\end{equation}
Substitution of equation~(\ref{NMsol}) into equation~(\ref{evnmom}) yields
\begin{equation}
\langle r^{2m}\rangle_s=2m\Gamma(2m)\frac{\wt{W}(s)^m}{s^{m+1}}\,.
\label{moms}
\end{equation}
For the moment of zero order, it is necessary to take into account that
$\lim_{m\to 0} m\Gamma(2m)=1/2$, therefore, $\langle r^{0}\rangle_s=1/s$,
i.e., the normalization condition.

For the variance ($m=1$), equation~(\ref{moms}) reads
\begin{equation}
\langle r^{2}\rangle_s=2\frac{\wt{W}(s)}{s^{2}}\,.
\label{varFP}
\end{equation}
The kernel of the time-nonlocal Fokker-Planck equation is defined by
the mean square displacement and, therefore, [see equation~(\ref{FPpdf})] the PDF
is also defined by this quantity.

\section{Determining the kernel from the Langevin equation}
\label{Langevin}

The conclusion of the last section is that in order to obtain the appropriate
Fokker-Planck equation for a given process, we need to determine the time dependence
of the variance $\langle r^2 \rangle_t$. One way to do so is to measure this moment from experimental
data. On the other hand, if this data are not available, or if one wants to derive this information
from the physics of the problem, another starting point can be the generalized Langevin
equation~\cite{L908,LG97}.

The standard Langevin equation is  Newton's second law applied to a Brownian
particle where the random force acting on a particle is taken into account.
As was shown in~\cite{M65,KI66,K66}, the generalized Langevin
equation is written in terms of a time non-local friction
force:
\begin{equation}
\frac{\partial^2 r(t)}{\partial t^2}=-\int\limits_{0}^{t}
\gamma(t-t^\prime) \frac{\partial r(t^\prime)}{\partial t^\prime}\ud t^\prime+
R(t),
\label{lang2}
\end{equation}
where $R(t)$ is the random force component which is uncorrelated
with the velocity that has a zero mean
$\langle R(t)\rangle=0$.
The autocorrelation of the random force is related to the kernel in
equation~(\ref{lang2}) by the fluctuation-dissipation theorem~\cite{KI66,K66}:
\begin{equation}
\langle R(t)R(t^\prime)\rangle=\frac{k_\mathrm{B}T}{m} \gamma(t-t^\prime),
\label{fd2}
\end{equation}
where $m$ is the mass of a particle.
For the kernel of the kind $\gamma(t)=\gamma_0\delta(t)$, equation~(\ref{lang2}) is
transformed to the standard Langevin equation.
In the general case, the Fourier transform of the
random force correlation function is
colored, for example cf.~\cite{L05}).

In dimensionless variables defined in the section \ref{telegraph}
with the diffusion coefficient defined by $D=k_\mathrm{B} T t_\mathrm{c}/m$ and
$1/t_\mathrm{c}=\int_0^{\infty} \gamma(t) \ud t$,
the Laplace transform of the mean square displacement follows from
equation~(\ref{lang2}) (see, e.g.,~\cite{PWM96}) and reads
\begin{equation}
{\widetilde{\langle \xi^2\rangle}_{s}}=
\frac{2}{s^2\left[s+\tilde{\gamma}(s)\right]}\,.
\label{xHlap}
\end{equation}
where the dimensionless kernel  satisfies in the time domain
$\int_0^{\infty}\gamma(\tau)\ud \tau=1$.
One can see from this equation that if
 $\lim_{s\to\infty}\tilde{\gamma}(s)/s\to 0$,  the function
$\widetilde{\langle \xi^2\rangle}_{s}\sim 1/s^3$
and the mean square displacement exhibit the ballistic behavior at short times
$\widetilde{\langle \xi^2\rangle}_{t}\sim t^2$. The shape of the function
$\tilde{\gamma}(s)$ at small $s$ is responsible for asymptotic behavior of the
mean square displacement in the time domain.

Substitution of equation~(\ref{xHlap}) into equation~(\ref{varFP}) yields the
kernel
\begin{equation}
\wt{W}(s)=\frac{1}{s+\tilde{\gamma}(s)}\,.
\label{Lkernel}
\end{equation}
This equation settles the relation between the memory kernel of the
time-nonlocal Fokker-Planck equation and the memory kernel of the Langevin
equation.

The solution of equation~(\ref{FPpdf}) given by equation~(\ref{NMsol}) with the memory
kernel defined by equation~(\ref{Lkernel}) reads
\begin{equation}
\wt{f}(\xi,s)=\sqrt{\frac{2}{\pi}}\xi\left[s+\tilde{\gamma}(s)\right]
\frac{K_{1/2}\left(\xi\sqrt{s\left[s+\tilde{\gamma}(s)\right]}\right)}
{\left\{\xi\sqrt{s\left[s+\tilde{\gamma}(s)\right]}\right\}^{1/2}}\,.
\label{solLkern}
\end{equation}
Equation~(\ref{solLkern}) consists of two terms,
\begin{equation}
\wt{f}(\xi,s)=\xi\left[s+\tilde{\gamma}(s)\right]\wt{\Psi}\left(\xi^2s\left[s+\tilde{\gamma}(s)\right]\right),
\label{solPs}
\end{equation}
where
\begin{equation}
\wt{\Psi}(z)=\sqrt{\frac{2}{\pi}}\frac{K_{1/2}\left(\sqrt{z}\right)}{\left(\sqrt{z}\right)^{1/2}}.
\label{Psi}
\end{equation}
The Taylor expansion of equation~(\ref{Psi}) is defined by
\begin{equation}
\wt{\Psi}(z)=\sum\limits_{n=0}^\infty\frac{1}{n!}
\frac{\partial ^n}{\partial z^n}\wt{\Psi}(z)\Bigg|_{z=z_0}(z-z_0)^n\,.
\label{Texp}
\end{equation}
For the modified Bessel function of the third kind
\begin{equation}
\frac{\partial ^n}{\partial z^n}
\frac{K_{1/2}\left(\sqrt{z}\right)}{\left(\sqrt{z}\right)^{1/2}}=
\frac{(-1)^n}{2^n}\frac{K_{1/2+n}\left(\sqrt{z}\right)}{\left(\sqrt{z}\right)^{1/2+n}}\,.
\label{derB}
\end{equation}
Equation~(\ref{solPs}) defines that $z=\xi^2s\left[s+\tilde{\gamma}(s)\right]=
z_0^2-\xi^2\left[\tilde{\gamma}(s)/2\right]^2$, where
$z_0=\xi^2\left\{s+\left[\tilde{\gamma}(s)/2\right]\right\}^2$, therefore, equation~(\ref{Texp}) reads
\begin{equation}
\wt{\Psi}(z)=\sqrt{\frac{2}{\pi}}\sum\limits_{n=0}^\infty\frac{1}{n!}
\frac{K_{1/2+n}\left(\xi\left[s+\tilde{\gamma}(s)/2\right]\right)}{\left\{\xi\left[s+\tilde{\gamma}(s)/2\right]\right\}^{1/2+n}}
\left[\frac{\xi\tilde{\gamma}(s)}{2\sqrt{2}}\right]^{2n}\,.
\label{solPsiser}
\end{equation}
It is suitable to rewrite equation~(\ref{solPs}) in the following form
\begin{eqnarray}
\wt{f}(\xi,s)&=&r\left\{\left[s+\tilde{\gamma}(s)/2\right]+\tilde{\gamma}(s)/2\right\}
\wt{\Psi}\left(r^2s\left[s+\tilde{\gamma}(s)\right]\right)\nonumber \\
&=&\wt{f}_{I}(\xi,s)+\wt{f}_{II}(\xi,s)\,.
\label{solPs1}
\end{eqnarray}
Taking into account that
\begin{equation}
K_{1/2}(z)=\sqrt{\frac{\pi}{2}}\frac{\re^{-z}}{\sqrt{z}}
\label{K12}
\end{equation}
equation~(\ref{solPsiser}) can be written as
\begin{eqnarray}
&&\wt{\Psi}(z)=\frac{\exp\left\{-\xi\left[s+\tilde{\gamma}(s)/2\right]\right\}}{\xi\left[s+\tilde{\gamma}(s)/2\right]}
+
\sum\limits_{n=1}^\infty\frac{1}{n!}
\frac{K_{1/2+n}\left(\xi\left[s+\tilde{\gamma}(s)/2\right]\right)}{\left\{\xi\left[s+\tilde{\gamma}(s)/2\right]\right\}^{1/2+n}}
\left[\frac{\xi\tilde{\gamma}(s)}{2\sqrt{2}}\right]^{2n}\,.
\label{extrsolv}
\end{eqnarray}
Let $\lim_{s\to\infty}\tilde{\gamma}(s) =\gamma_0$. Under this assumption, the
sum in
equation~(\ref{extrsolv}) asymptotically tends to zero and the function defined
by equation~(\ref{extrsolv}) reads
\begin{equation}
\wt{\Psi}(z)\Bigg|_{s\to\infty}\sim
\frac{\exp\left[-\xi\left(s+{\gamma}_0/2\right)\right]}{\xi\left(s+{\gamma}_0/2\right)}
\label{PsiAs}
\end{equation}
and the function $\wt{f}_{II}(\xi,s)$ from equation~(\ref{solPs1}) is given by
\begin{equation}
\wt{f}_{II}(\xi,s)\Bigg|_{s\to\infty}\sim
\gamma_0\frac{\exp\left[-\xi\left(s+{\gamma}_0/2\right)\right]}{s+{\gamma}_0/2}\,.
\label{f2s}
\end{equation}
The inverse Laplace transform of equation~(\ref{f2s}) yields
\begin{equation}
{f}_{II}(\xi,t)=\gamma_0\re^{-{\gamma_0}\tau/{2}}\Theta(\tau-\xi).
\label{f2t}
\end{equation}
proceeding to limit $\lim_{\tau\to 0}{f}_{II}(\xi,\tau)=0$ shows that this
contribution to the PDF has nothing to do with the initial condition and is
responsible for diffusion  of injecting particles during the transport
process.

The nonvanishing term in the function $\wt{f}_{I}(\tau,s)$ in the limit of
large $s$ is defined by
\begin{equation}
\wt{f}_{I}(\xi,s)\Bigg|_{s\to\infty}\sim
\exp\left[-\xi\left(s+{\gamma}_0/2\right)\right].
\label{f1s}
\end{equation}
Its inverse Laplace transform yields
\begin{equation}
{f}_{I}(\xi,t)=\re^{-{\gamma_0}\tau/{2}}\delta(\tau-\xi).
\label{f1t}
\end{equation}
Therefore, the part of the PDF under consideration which is responsible for
the initial condition and the further impulse propagation is contained in the
first summand in equation~(\ref{solPs1}).

Now we can isolate from the PDF defined by equation~(\ref{solLkern}) the
part corresponding to the diffusion process of the particles which
are lost by the propagating impulse
\begin{eqnarray}
\wt{f}_\mathrm{diff}(\xi,s)&=&\wt{f}(\xi,s)-\exp\left[-\xi\left(s+{\gamma}_0/2\right)\right]\nonumber \\
&=&\wt{\Phi}(\xi,s)\re^{-\xi s},
\label{smPDF}
\end{eqnarray}
where the function $\wt{\Phi}(r,s)$ is defined by equation~(\ref{solLkern}) and
equation~(\ref{K12}) and reads
\begin{eqnarray}
\wt{\Phi}(\xi,s)&=&\left[s+\tilde{\gamma}(s)\right]
\frac{\exp\left(-\xi\left\{\sqrt{s\left[s+\tilde{\gamma}(s)\right]}-s\right\}\right)}
{\sqrt{s\left[s+\tilde{\gamma}(s)\right]}}
-\exp\left(-\frac{\xi\gamma_0}{2}\right).
\label{Phsm}
\end{eqnarray}
The inverse Laplace transform of equation~(\ref{Phsm}) yields the diffusive part
of the PDF in the time domain
\begin{equation}
{f}_\mathrm{diff}(\xi,t)=\Phi(\xi,\tau-\xi)\Theta(\tau-\xi).
\label{PhsmT}
\end{equation}
The importance of this result is that the explicit Heaviside function
takes upon itself the discontinuity in the function $f_\mathrm{diff}(\xi,\tau)$.
The function $\Phi(\xi,\tau-\xi)$ in the time domain is a continuous
function. If it does not have an analytical representation in a closed form,
it can be evaluated numerically, for example using the direct
integration method~\cite{duf93}.

Summing together the results (\ref{f1t}) and (\ref{PhsmT}) in the time domain
we get a general solution of the non-Markovian problem with a short-time
behavior, in the form
\begin{eqnarray}
f(\xi,\tau)&=&\re^{-{\gamma_0}\tau/{2}}\delta(\tau-\xi)+\Phi(\xi,\tau-\xi)\Theta(\tau-\xi).
\label{fullsolb}
\end{eqnarray}
From this solution one can see  that the diffusion repartition of the PDF
occurs inside the spatial diffusion domain which increases in a deterministic
way. The first term in equation~(\ref{fullsolb}) corresponds to the propagating
delta-function which is inherited from the initial conditions, and it keeps
decreasing in time at the edge of the ballistically expanding domain.

In order to estimate the PDF in the long time limit, it is necessary to
suggest the form of  the memory kernel of the Langevin equation at small $s$.
The reasonable choice is given by
\begin{equation}
\wt{\gamma}(s)_{s\sim 0}\sim  s^{\alpha -1}.
\label{GsmallS}
\end{equation}
One can see from equation~(\ref{xHlap}) that under  condition $s^{\alpha+1}\geqslant s^3$
(i.e., $\alpha\leqslant 2$) the inverse Laplace transform of equation~(\ref{xHlap})
yields the variance in the form of equation~(\ref{var2}).
Substitution of equation~(\ref{GsmallS}) to equation~(\ref{solLkern}) defines the
Laplace transform of the PDF
at small $s$
\begin{equation}
\wt{f}(\xi,s)=\frac{\exp\left(-\xi s^{\alpha /2}\right)}{s^{1-\alpha /2}}\,.
\label{PDFsmallS}
\end{equation}
In the general case, the inverse Laplace transform of equation~(\ref{PDFsmallS})
is given by the Fox  functions (see, e.g.~\cite{MK00}).
For $\alpha =1$ in the time domain, the PDF is defined by equation~(\ref{classol}).
For the special case $\alpha=0$, the inverse Laplace transform of
equation~(\ref{PDFsmallS}) is independent of time and reads
\begin{equation}
f(\xi)=\re^{-\xi}.
\label{PDFa0}
\end{equation}
This result coincides with the PDF from~\cite{BHW87}.

Below we demonstrate the application of the time-nonlocal approach
with explicit examples.

\subsection{Standard asymptotic diffusion}

For $\wt{\gamma}(s)=1$ in equation~(\ref{Lkernel})
the generalized Langevin
equation is reduced to the standard one which predicts the mean square
displacement in the form of equation~(\ref{vartlg}).
The memory kernel of the time-nonlocal Fokker-Planck equation reads
\begin{equation}
\wt{W}(s)=\frac{1}{s+1}\,.
\label{FPclkern}
\end{equation}
In this case, equation~(\ref{smPDF}) reads
\begin{equation}
\wt{f}_\mathrm{diff}(\xi,s)=\wt{\Phi}(\xi,u)\re^{-\xi u},
\label{smPDFl}
\end{equation}
where $u=s+1/2$, therefore, the inverse Laplace transform is given by
\begin{equation}
f_\mathrm{diff}(\xi,t)=\re^{-\frac{1}{2}t}\Phi(\xi,t-\xi)\Theta(t-\xi).
\label{solBsm}
\end{equation}
The function $\wt{\Phi}(\xi,u)$ for the particular case under consideration
follows from the definition given by equation~(\ref{Phsm})
\begin{equation}
\wt{\Phi}(\xi,u)=u\Psi(\xi,u)-1+\frac{1}{2}\Psi(\xi,u),
\label{philc}
\end{equation}
where
\begin{equation}
\wt{\Psi}(\xi,u)=
\frac{\exp\left[-\mid \xi\mid\left(\sqrt{u^2-1/4}-u\right)\right]}
{\sqrt{u^2-1/4}}\,.
\label{fpru}
\end{equation}
From the initial value theorem, it follows that
\begin{eqnarray}
\Psi(\xi,t=0)&=&\lim_{u\to \infty}u\wt{\Psi}(\xi,u)
\nonumber \\
&=&\lim_{u\to \infty}\frac{\exp\left\{-\mid \xi\mid \left[u\sqrt{1-1/(4u^2)}-u\right]\right\}}
{\sqrt{1-1/(4u^2)}}=1,
\label{ivt}
\end{eqnarray}
Therefore, the inverse Laplace transform of equation~(\ref{philc}) is given by
\begin{equation}
\Phi(\xi,t)=\frac{\partial}{\partial t}\Psi(\xi,t)+\frac{1}{2}\Psi(\xi,t).
\label{pssi1}
\end{equation}
The inverse Laplace transform of equation~(\ref{fpru}) is given by~\cite{AS}
\begin{equation}
\Psi(\xi,t)=I_{0}\left(\frac{1}{2}
\sqrt{t^2+2\xi t}\right),
\label{if2com1}
\end{equation}
where $I_0(z)$ is the modified Bessel function.

Substitution of equation~(\ref{if2com1}) into equation~(\ref{pssi1}) yields
\begin{eqnarray}
\Phi(\xi,t)=\frac{1}{2}\left[I_0\left(\frac{1}{2}\sqrt{t^2+2\xi t}\right)
+(t+\xi)\frac{I_1\left(\frac{1}{2}\sqrt{t^2+2\xi t}\right)}{\sqrt{t^2+2\xi t}}\right].
\label{inv1}
\end{eqnarray}
Substitution of this equation into equation~(\ref{fullsolb}) reduces the solution to
the result for the telegraph equation given by equation~(\ref{soltlg}).

\begin{figure}[!b]
\centerline{
\includegraphics[width=0.45\textwidth]{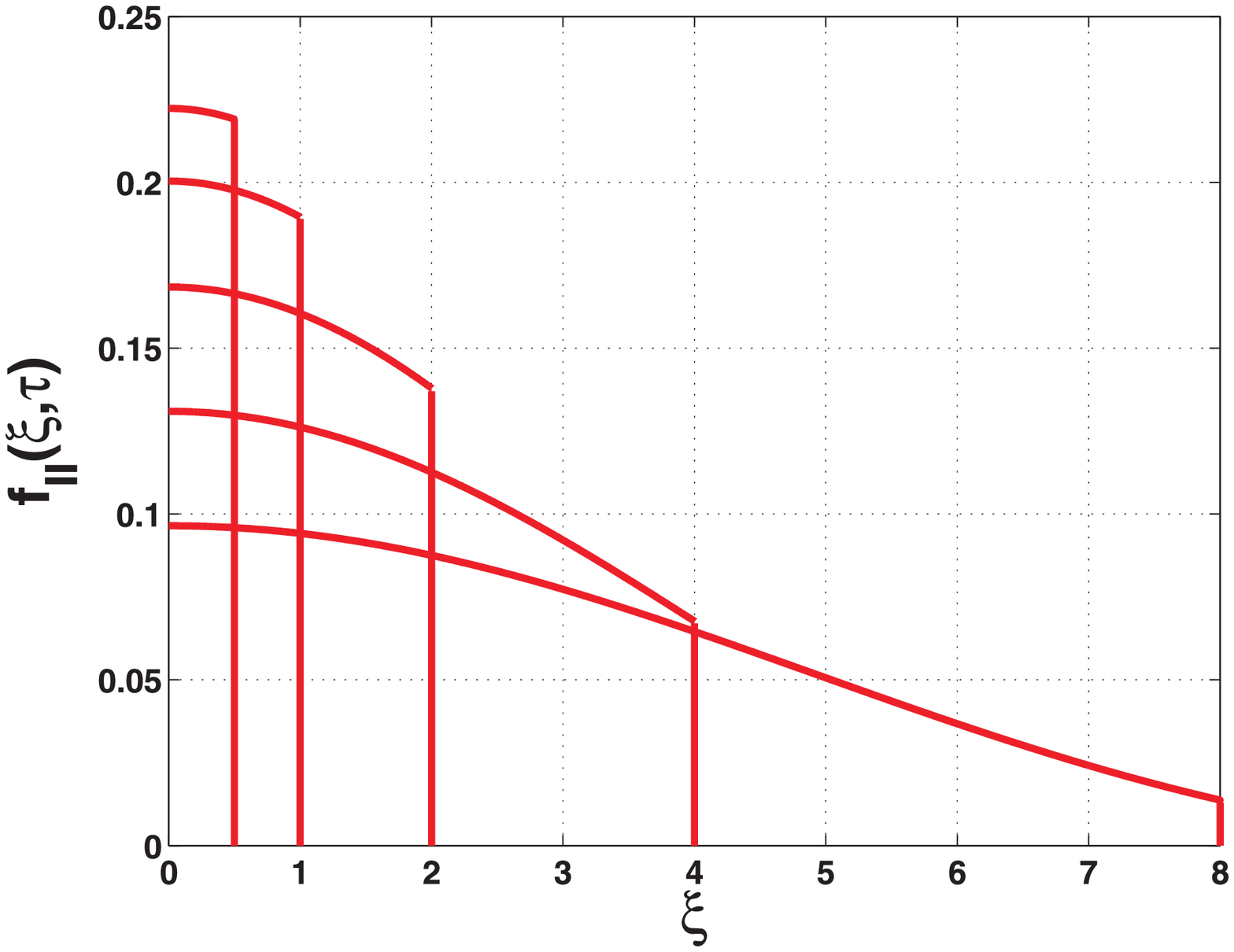}
\hspace{1cm}
\includegraphics[width=0.435\textwidth]{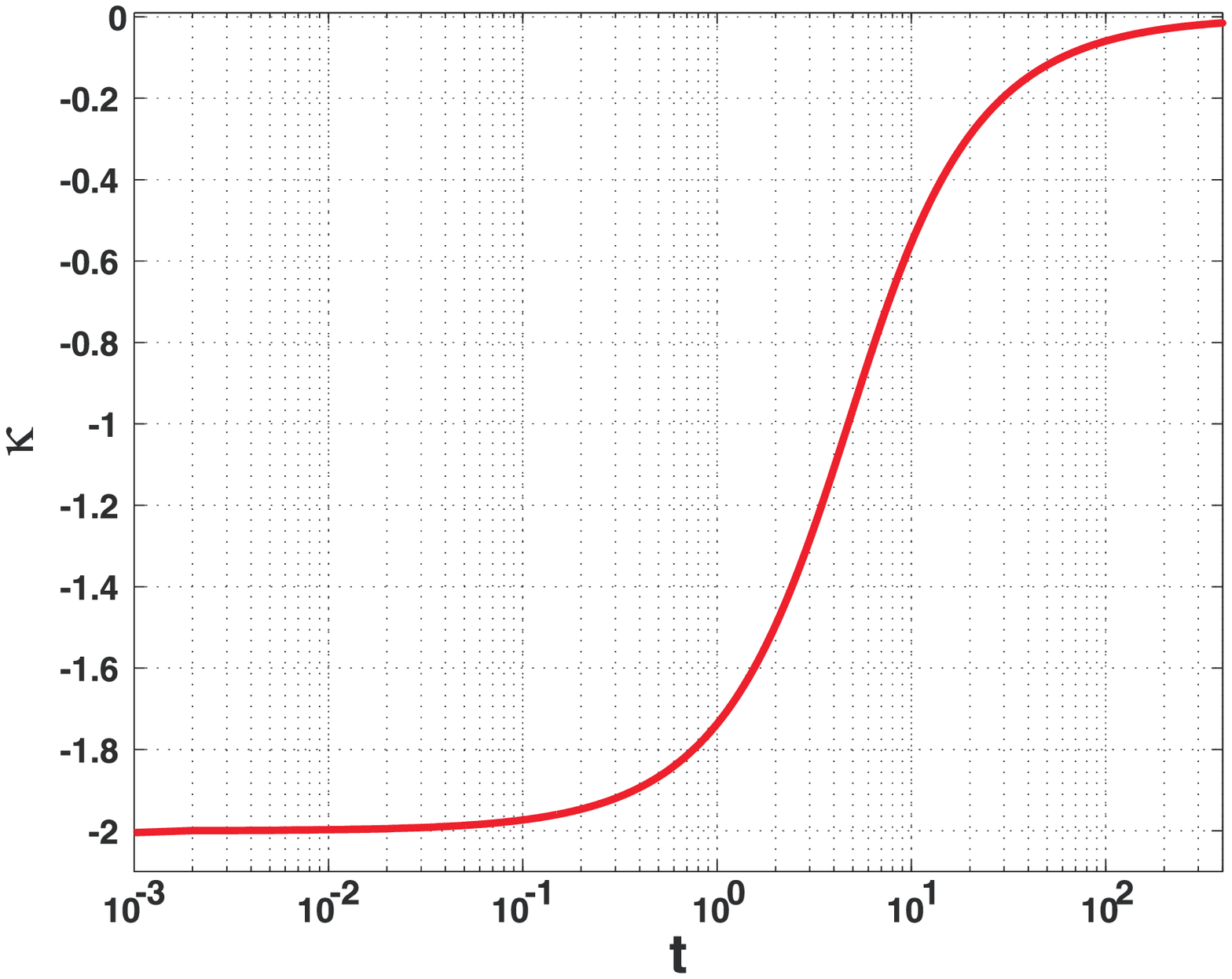}
}
\vspace{-2mm}
\parbox[t]{0.5\textwidth}{%
\caption{(Color online) The diffusion  part of the PDF defined by equation~(\ref{soltlg})
for time intervals $\tau=$0.5, 1, 2, 4, 8.}
\label{fig2}}%
\parbox[t]{0.5\textwidth}{%
\caption{(Color online) The time dependence of the kurtosis of the PDF corresponding
to the memory kernel defined by equation~(\ref{FPclkern}).}
\label{fig3}}
%
\end{figure}

The diffusion part of this solution [the sum of graphics shown in panels (b) and (c) in figure~\ref{fig1}] is displayed in figure~\ref{fig2}.
As was discussed above in the long time limit, this part approaches the
solution of the classical diffusion problem given by equation~(\ref{cldiff}).
In order to measure how fast the convergence takes place, it is convenient to estimate the
time dependence of the kurtosis of the distribution defined by
\begin{equation}
\kappa=\frac{\langle \xi^4\rangle}{\langle \xi^2\rangle^2}-3\,.
\label{kurt}
\end{equation}
For the Gauss distribution $\kappa=0$, for the pure ballistic propagation
$\kappa=-2$, moments are defined by equation~(\ref{moms}) and for the kernel
from equation~(\ref{FPclkern}) the second moment is given by equation~(\ref{vartlg}).
Calculation of the fourth moment yields
\begin{equation}
\langle \xi^4\rangle_\tau=12\left[\tau^2+6\left(1-\re^{-\tau}\right)-2\tau \left(3+\re^{-\tau}\right)\right].
\label{mom4}
\end{equation}

The kurtosis calculated with these moments is shown in figure~\ref{fig3}. At short
times, the propagation is ballistic with the following transition to the standard
diffusion during a large time interval.
\subsection{Anomalous diffusion}

\begin{figure}[!b]
\centering
\includegraphics[width=0.4\textwidth]{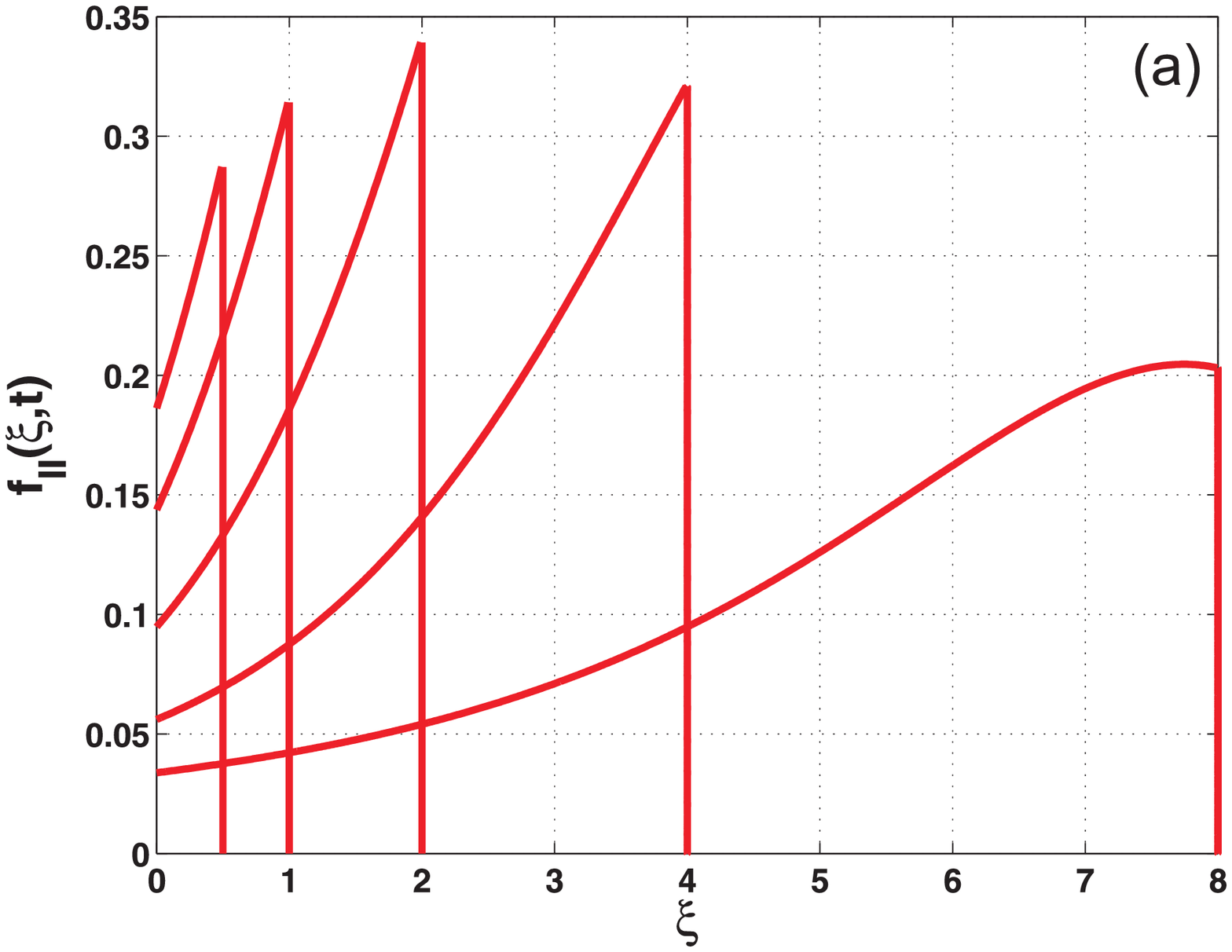}
\hspace{1cm}
\includegraphics[width=0.4\textwidth]{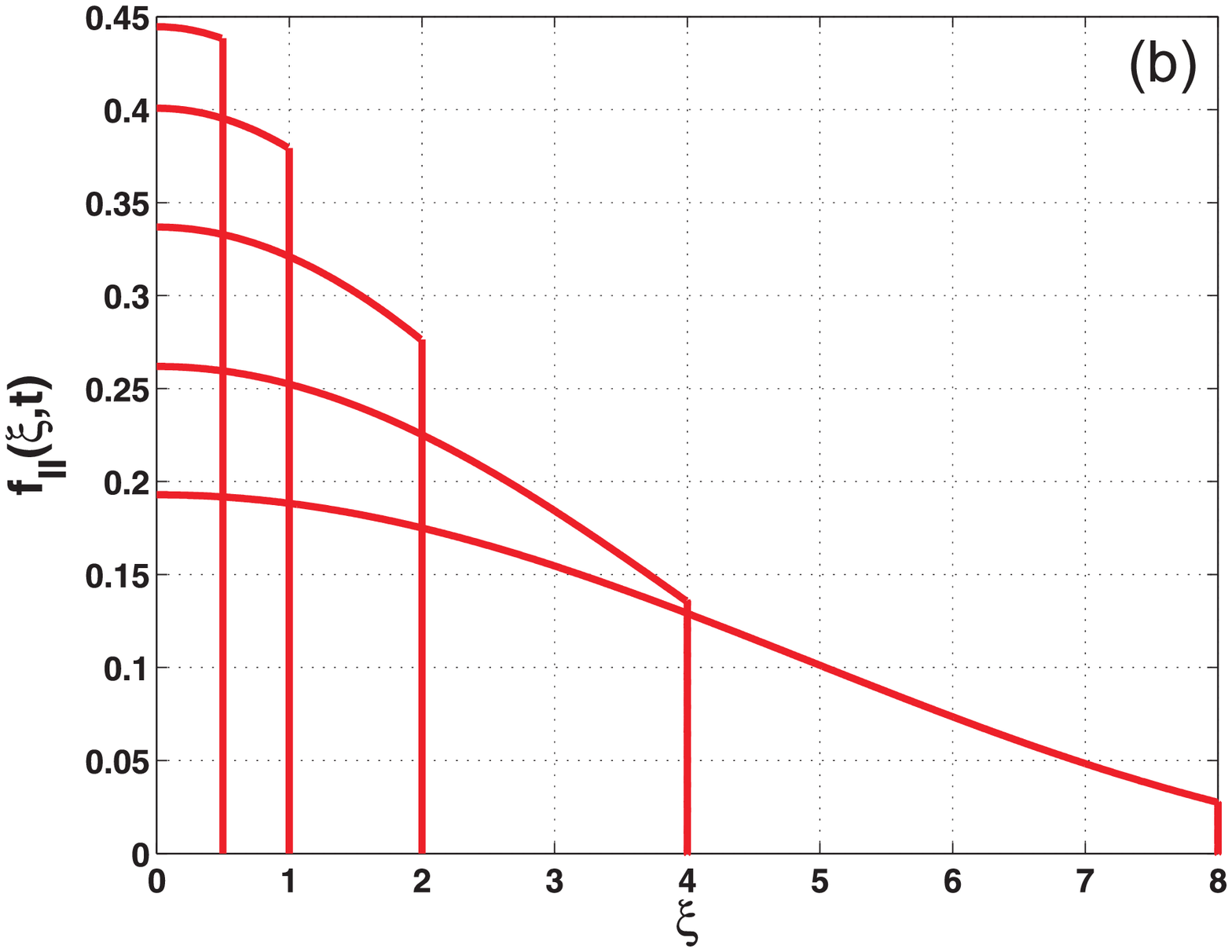}\\
\includegraphics[width=0.4\textwidth]{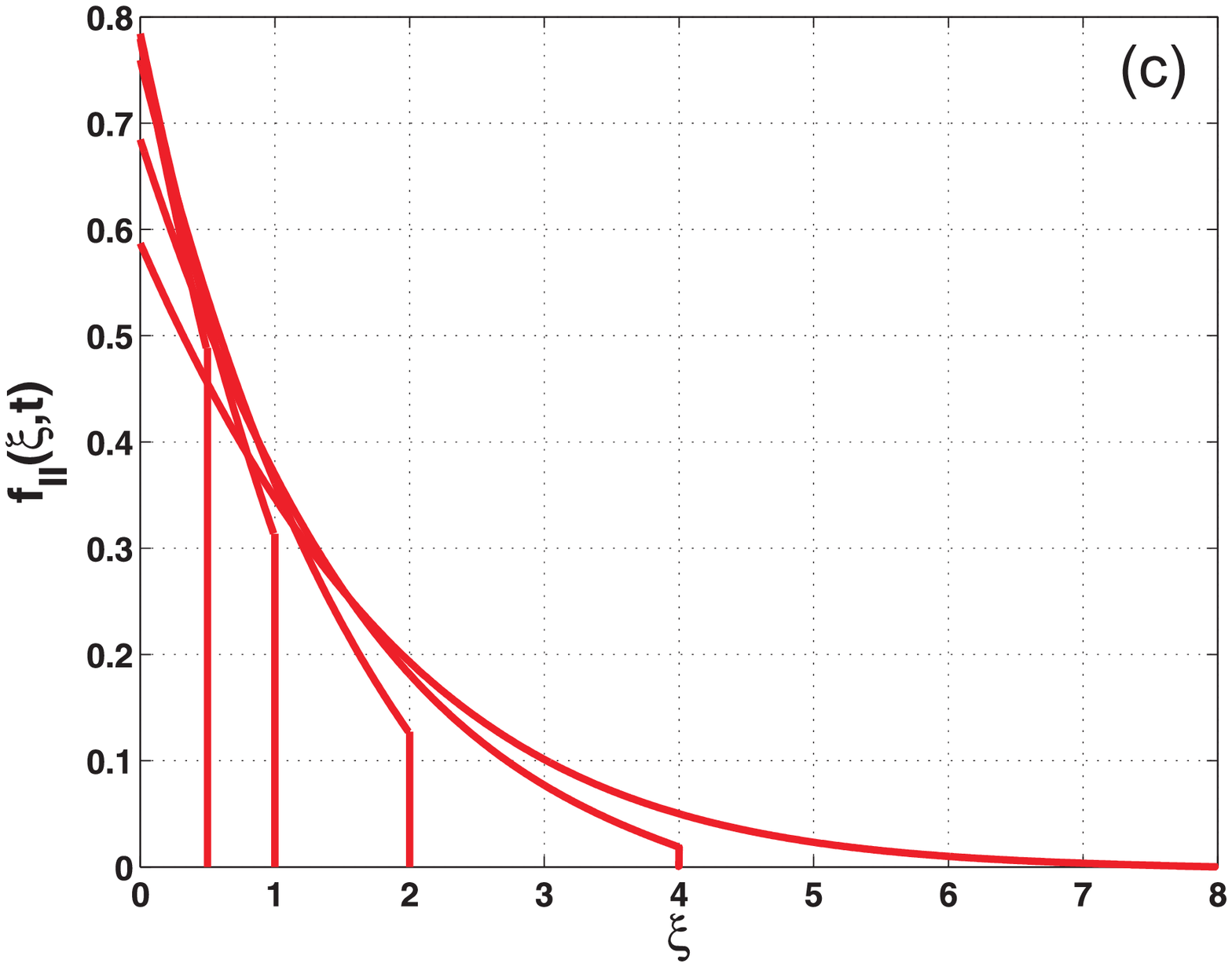}
\caption{(Color online) The diffusion part of the PDF defined by the memory kernel of the
Langevin equation [see equation~(\ref{LkernJ})] for different values of the
parameter $\alpha$. Superdiffusion
[$\alpha=3/2$, panel (a)], regular diffusion [$\alpha=1$, panel (b)] and
subdiffusion [$\alpha=1/2$, panel (c)].  Time intervals from the top to the
bottom $\tau=$0.5, 1, 2, 4, 8. }
\label{fig4}
\end{figure}

\begin{figure}[!t]
\centering
\includegraphics[width=0.45\textwidth]{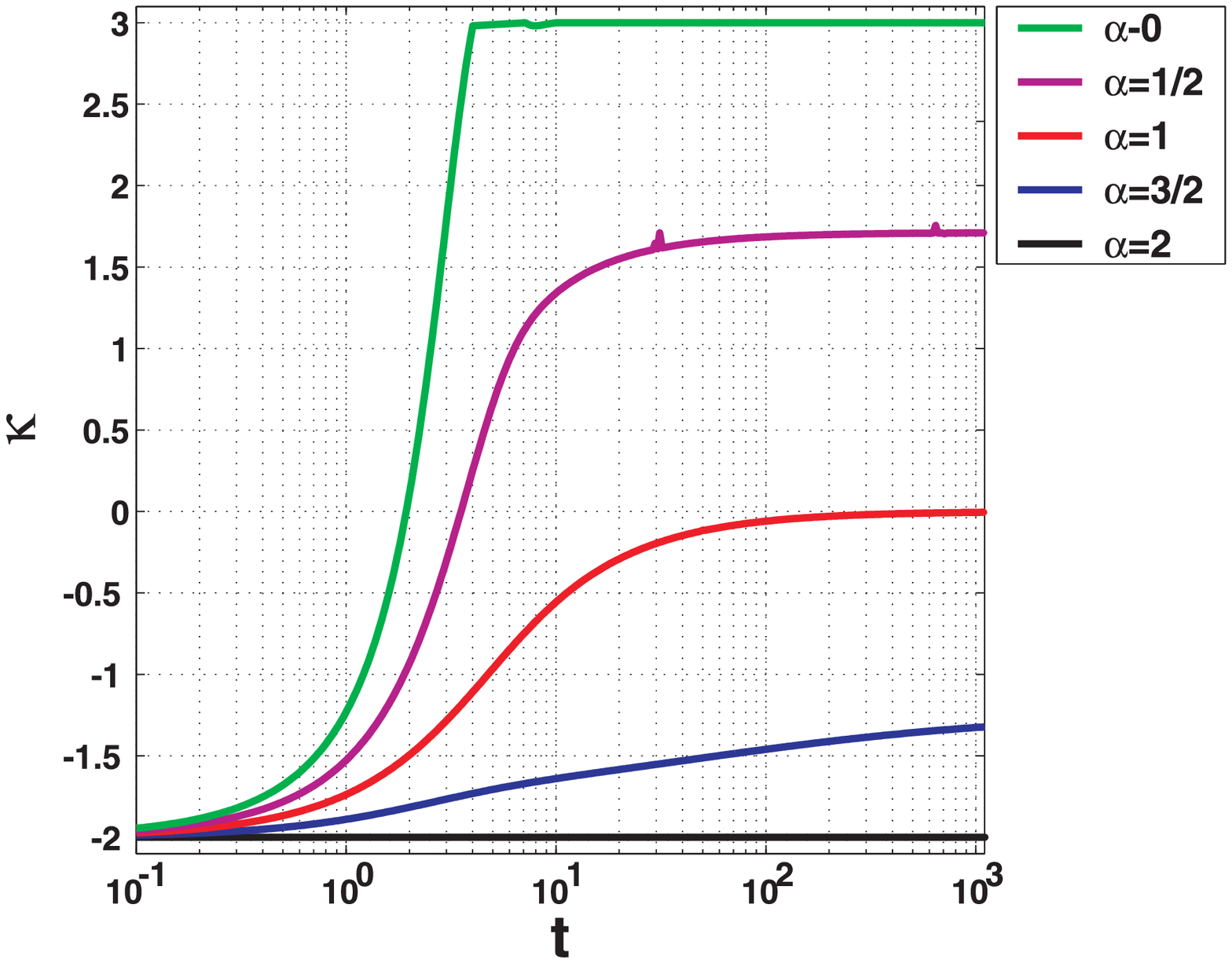}
\caption{(Color online) The time dependence of the kurtosis of the PDF corresponding
to the memory kernel defined by equation~(\ref{LkernJ}).}
\label{fig5}
\end{figure}

In order to join the short time ballistic and long time anomalous regimes
[see equation~(\ref{var2}) and equation~(\ref{GsmallS})], the Laplace transform of the
memory kernel of the Langevin equation can be defined by
\begin{equation}
\wt{\gamma}(s)=(2-\alpha)\frac{s^{\alpha-1}}{(1+s)^{\alpha-1}}\,.
\label{LkernJ}
\end{equation}
It follows from this equation that
\begin{equation}
\gamma_0=2-\alpha
\label{G0}
\end{equation}
and, therefore, the singular part of the PDF is given by
\begin{equation}
f_{I}(\xi,t)=\frac{1}{2}\exp\left[-\frac{3(2-\alpha)}{2}\tau\right]
\delta(\mid \xi \mid-\tau)\;.
\label{gf1}
\end{equation}
The diffusion part of the PDF was estimated numerically and is presented
in figure~\ref{fig4}. The corresponding kurtosis
 is shown in figure~\ref{fig5}.

\begin{figure}[!b]
\centering
\includegraphics[width=0.4\textwidth]{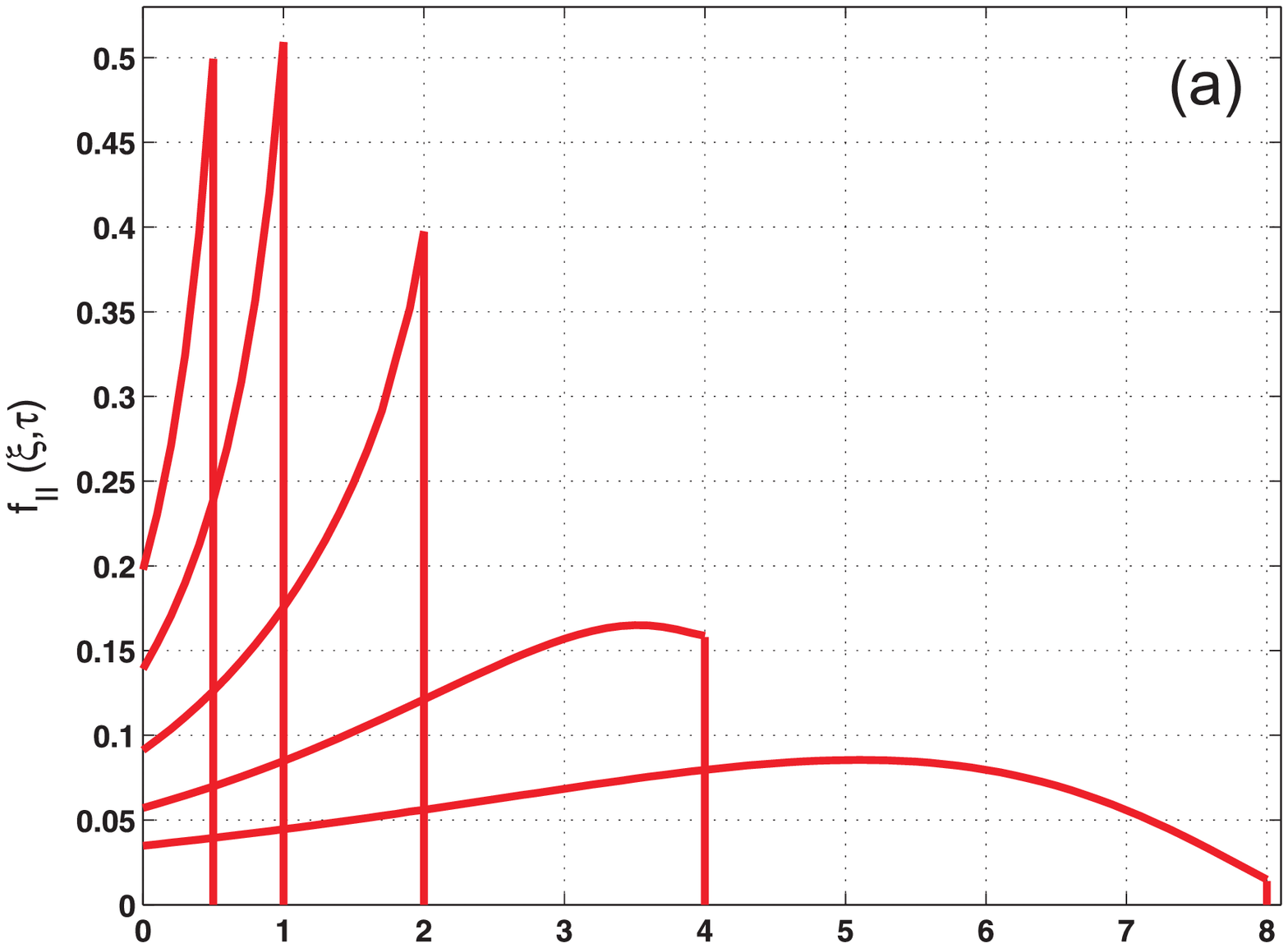}
\hspace{1cm}
\includegraphics[width=0.4\textwidth]{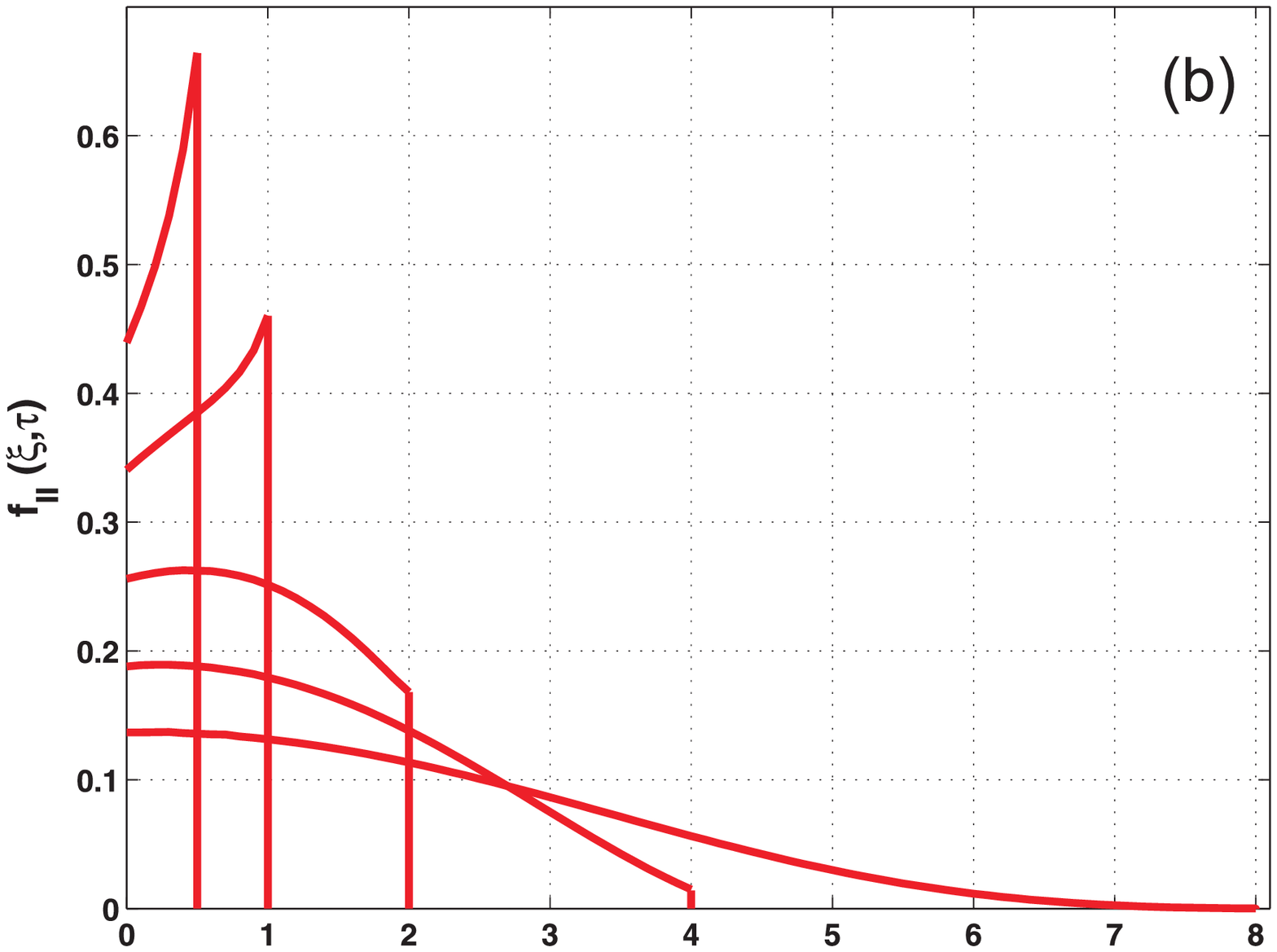}\\[2ex]
\includegraphics[width=0.4\textwidth]{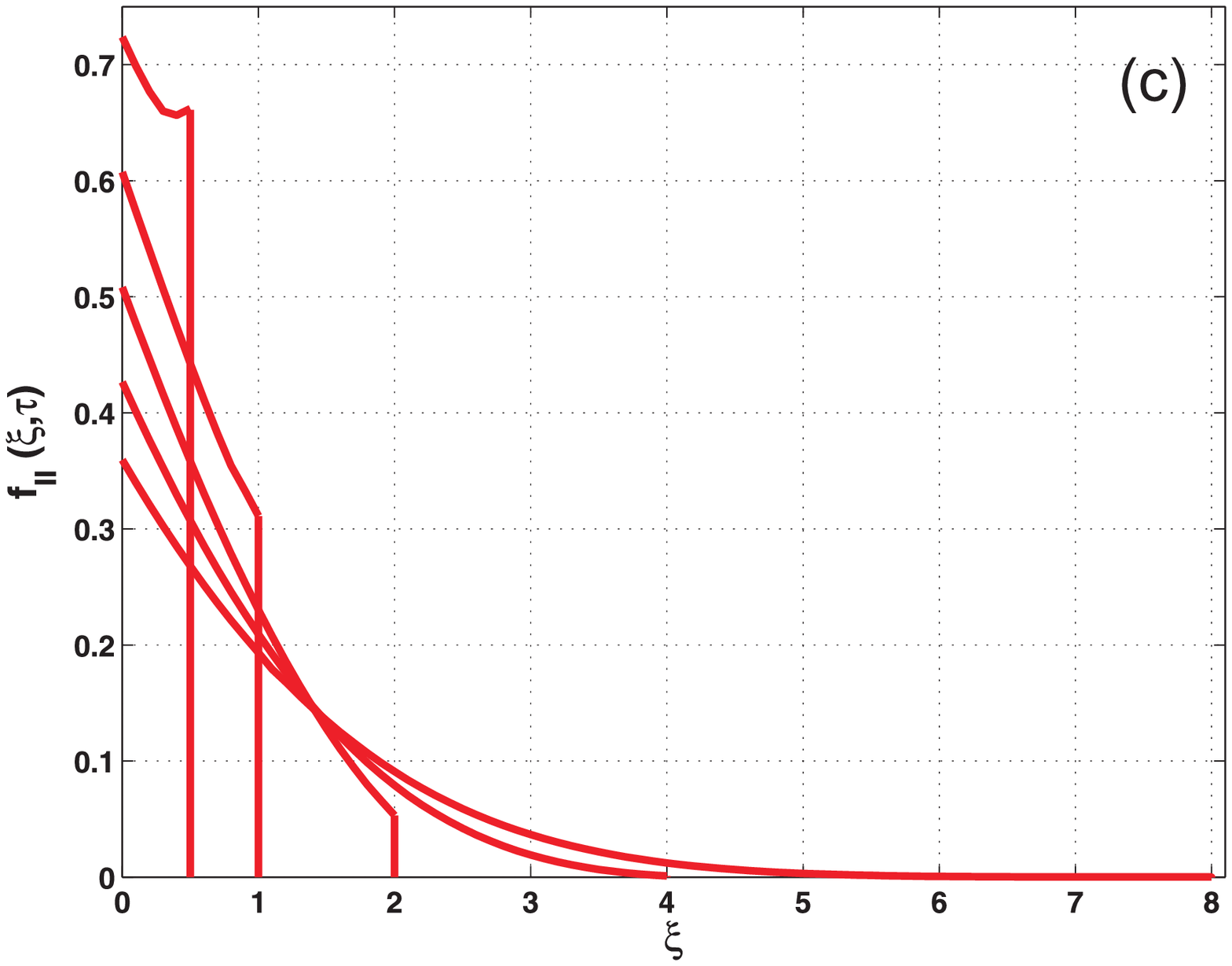}
\caption{(Color online) The diffusion part of the PDF corresponding to the mean square
displacement given by equation~(\ref{gdisp})  for different values of the
parameter $\alpha$. Superdiffusion
[$\alpha=3/2$, panel (a)], regular diffusion [$\alpha=1$, panel (b)] and
subdiffusion [$\alpha=1/2$, panel (c)].  Time intervals from the top to the
bottom $\tau=$0.5, 1, 2, 4, 8. }
\label{fig6}
\end{figure}

In a different way, the kernel of the time-nonlocal Fokker-Planck equation
can be defined by the mean square displacement given in the time domain.
In~\cite{IPZ10} this quantity was proposed in a form interpolating the
short time ballistic and long time anomalous behavior
\begin{equation}
\langle\Delta r^{2}\rangle_t=2D_{\alpha}t_{0}^{\alpha}\frac{(t/ t_{0})^{2}}
{\left[1+(t / t_{0})\right]^{2-\alpha}}\,,
\label{gdisp}
\end{equation}
where $0\leqslant \alpha\leqslant 2$ and  $t_{0}$ is the crossover characteristic time,
at $t\ll t_{0}$, the law (\ref{gdisp}) describes the ballistic regime and at
$t\gg t_{0}$ the fractional diffusion.

Introduce now dimensionless variables $\langle\xi^2\rangle _\tau =
\langle\Delta r^{2}\rangle _{t}/(2 D_{\alpha}t_{0}^{\alpha})$ and
$\tau=t/t_0$. With these variables, the last equation reads
\begin{equation}
\langle\xi^2\rangle _\tau =\frac{\tau^2}{(1+\tau)^{2-\alpha}}\,.
\label{gdispd}
\end{equation}

The Laplace transform of equation~(\ref{gdispd}) is given by
\begin{equation}
\widetilde{\langle \xi^2\rangle}_s=\left(\frac{\alpha}{s}-1\right)\frac{1}{s}+
\left[ (\alpha-1)\left(\frac{\alpha}{s}-2\right)+s\right]\frac{\re^s}{s^\alpha}
\Gamma(\alpha-1,s)\;,
\label{img}
\end{equation}
where $\Gamma(a,s)$ is the incomplete gamma function.
At large $s$ equation~can be written as
\begin{equation}
s^2\frac{\widetilde{\langle \xi^2\rangle}_{s}}{k_{B}T}\Bigg|_{s\to\infty}
\sim\frac{2}{s}-\frac{2\gamma_0}{s^2}+\ldots \, .
\label{r2large}
\end{equation}
Substitution of equation~(\ref{img}) into equation~(\ref{r2large}) yields the estimation
of the asymptotic value of the memory kernel $\gamma_0=3(2-\alpha)$ which
defines the time evolution of the singular part of the PDF. The results of
numerical calculations of the diffusive part are shown in figure~\ref{fig6}

\begin{figure}[!t]
\centering
\includegraphics[width=0.45\textwidth]{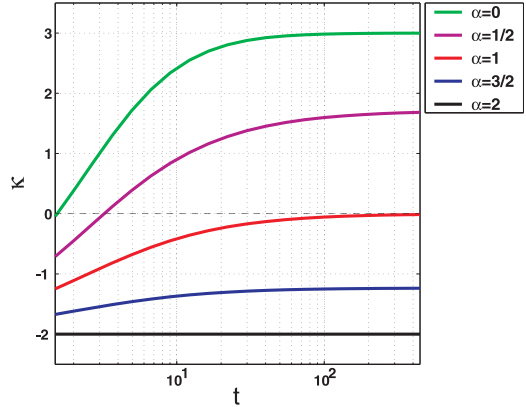}
\caption{(Color online) The time dependence of the kurtosis of the PDF corresponding
to the mean square displacement  defined by equation~(\ref{gdisp}).}
\label{fig7}
\end{figure}

The time evolution of the kurtosises of these distributions is displayed in
figure~(\ref{fig7}).

The reader should appreciate the tremendous role of
memory. The ballistic part which is represented by the advancing and reducing
delta-function sends backwards the probability that it loses due to the
exponential decay seen in figure~\ref{fig1}, panel (a). Thus, the diffusive
part of the PDF is replenished with time and asymptotically approaches
the Markovian distribution. The speed of convergence and the evolution of the
PDF shape depends on the interim behavior of the mean square displacement
(or, which is the same, of the memory kernel of the Langevin equation).

\section{Hydrodynamic theory of the time-nonlocal diffusion}
\label{hyd}
\subsection{Memory kernel}
The friction force acting upon a spherical particle has been derived in
\cite{B885,B61} (the modern discussion can be found in~\cite{LL87})  and is
given by
\begin{eqnarray}
F_\mathrm{fr}(t)=-2\pi R^3\left[\frac{3\eta}{R^2}u(t)+\frac{\rho}{3}
\frac{\partial u(t)} {\partial t} 
+\frac{3}{R}\sqrt{\frac{\rho\eta}{\pi}}
\int\limits_0^t \frac{1}{\sqrt{t-\tau}}\frac{\partial u(\tau)}{\partial \tau}
\ud\tau\right],
\label{FrBB}
\end{eqnarray}
where $F_\mathrm{fr}(t)$ is the friction force, $R$ is the particle radius,
and $\rho$ and $\eta$ are the density and viscosity of the solvent.
Substitution of this force into the Langevin equation yields the fractional
equation,  and the Laplace transform of the memory kernel
can be found in a straightforward way~\cite{MP96}. For simplicity
instead of this approach we will use the results obtained~\cite{VT45} for the mean square
displacement.

\begin{figure}[!h]
\centering
\includegraphics[width=0.45\textwidth]{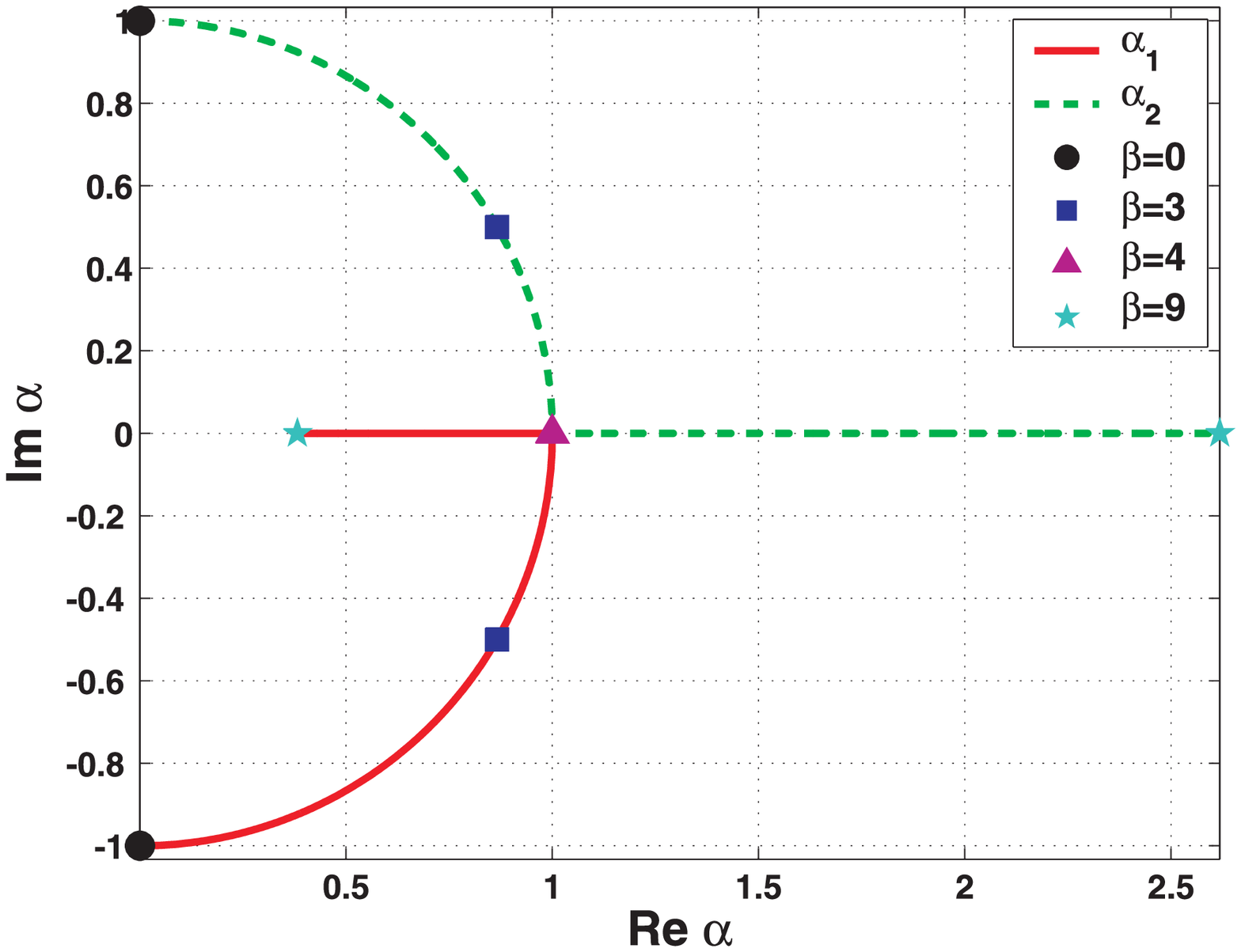}
\hspace{1cm}
\includegraphics[width=0.45\textwidth]{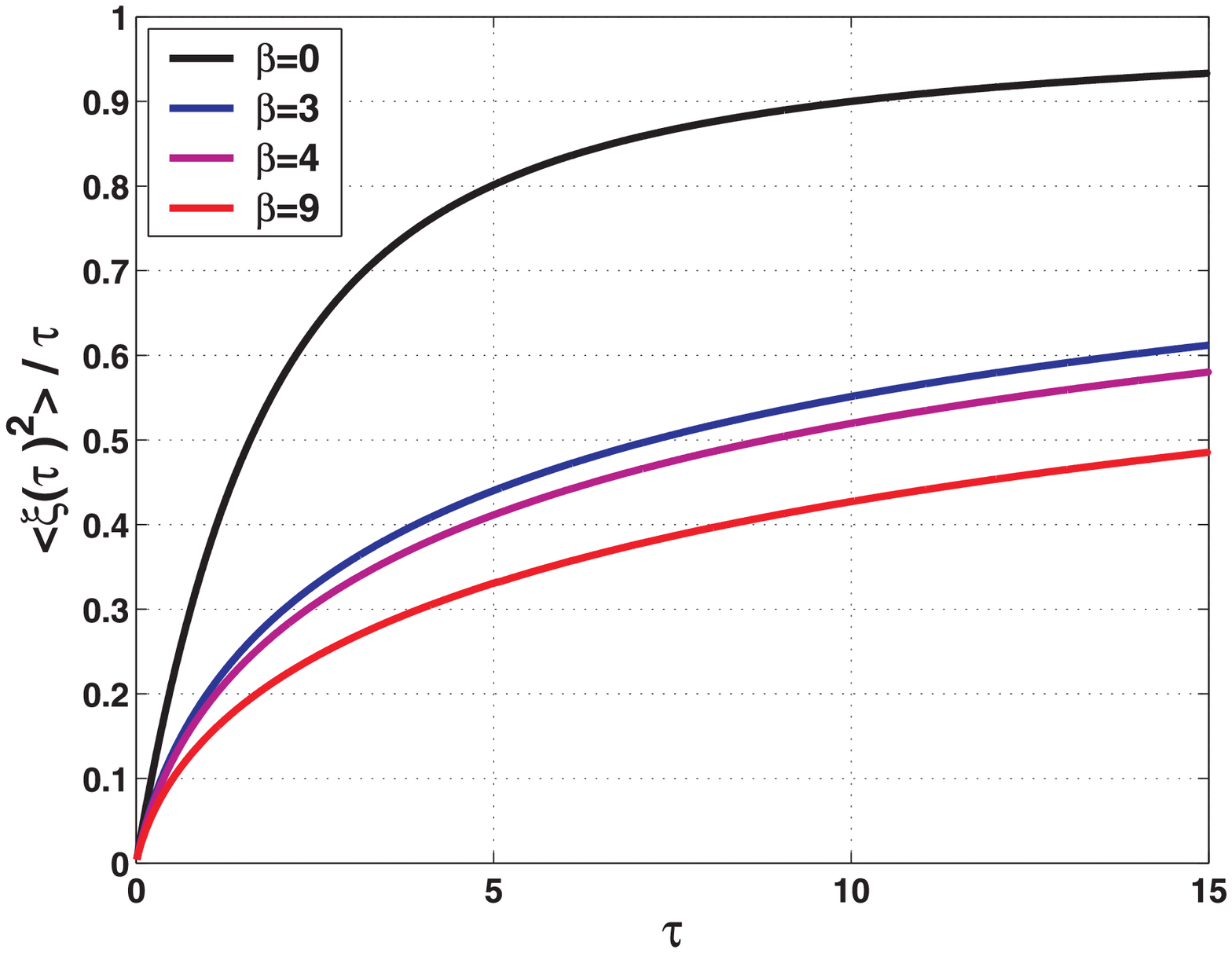}
\parbox[t]{0.49\textwidth}{%
\caption{(Color online) Dependence of coefficients $\alpha_{1,2}$ on the parameter $\beta$.\label{fig8}}
}
\hfill%
\parbox[t]{0.49\textwidth}{%
\caption{(Color online) Time dependence of the mean square displacement for different
values of the parameter $\beta$ [$\langle \xi(\tau )^2\rangle=
\langle x^2(\tau)\rangle/(2D\tau_\mathrm{F})$].\label{fig9}}
}
\end{figure}

A system with hydrodynamic memory has two characteristic times
\begin{equation}
\tau_R=\frac{9}{2}\frac{M_\mathrm{s}}{S}\,,
\label{tR}
\end{equation}
where $M_\mathrm{s}$ is the mass of a solvent particle, $S=6\pi \eta R$ is the
Stokes friction coefficient and
\begin{eqnarray}
\tau_\mathrm{F}=\frac{M+M_\mathrm{s}/2}{S} =\frac{1}{9}\Bigg(1+2\frac{M}{M_\mathrm{s}}\Bigg)\tau_R\,.
\label{tF}
\end{eqnarray}
One can see from equation~(\ref{tF}) that $\tau_\mathrm{F}>0$. Therefore, it is reasonable
to introduce  dimensionless variables $\tau=t/\tau_\mathrm{F}$ and
$\beta=\tau_R/\tau_\mathrm{F}$.
Then, the projection of the mean square displacement onto one of the axis reads
\begin{eqnarray}
\langle \Delta x^2(\tau)\rangle & = & 2D\tau_\mathrm{F}\Bigg\{\tau-
2\left({{\beta}\tau/{\pi}}\right)^{1/2}+(\beta-1) \nonumber \\
&&+\frac{1}{\alpha_1-\alpha_2}\Bigg[\frac{1}{\alpha_1^3}\re^{\alpha_1^2 \tau}
\mathrm{erfc}\left(\alpha_1 \sqrt{\tau}\right) -\frac{1}{\alpha_2^3}\re^{\alpha_2^2 \tau}
\mathrm{erfc}\left(\alpha_2 \sqrt{\tau}\right)\Bigg]\Bigg\}\;,
\label{msd1}
\end{eqnarray}
where   $D$ is the diffusion coefficient, $\mathrm{erfc}(z)$ is the complimentary
error function
and coefficients $\alpha_{1}$ and $\alpha_{2}$    are given by
\begin{equation}
\alpha_{1,2}=\frac{1}{2}\Bigg(\sqrt{\beta}\mp \sqrt{\beta-4}\Bigg).
\label{al}
\end{equation}
The dependence of the coefficients $\alpha_1$ and $\alpha_2$ on the parameter $\beta$
is shown in figure~\ref{fig8}.

At short times $\tau\ll 1$, the mean square displacement is given by the
Taylor
expansion of equation~(\ref{msd1})
\begin{eqnarray}
\langle \Delta x^2(\tau)\rangle &=& 2D\tau_\mathrm{F}\tau^2\Bigg\{\frac{1}{2}-
\frac{8}{15}\left({{\beta\tau}/{\pi}}\right)^{1/2}
+\frac{1}{6}(\beta-1)\tau \nonumber \\
&&+\frac{16}{105}\left({{\beta\tau}/{\pi}}\right)^{1/2}\tau
+\frac{1}{24}\left[\beta(\beta-3)+1\right]\tau^2+\cdots \Bigg\}.
\label{msds}
\end{eqnarray}
One can see that the mean square displacement defined by equation~(\ref{msds})
corresponds to the ballistic motion.

In the long time limit, the mean square displacement is given by
\begin{eqnarray}
\langle \Delta x^2(\tau)\rangle=2D\tau_\mathrm{F}\tau\left\{1-
2\left({{\beta}/{\pi\tau}}\right)^{1/2}+\frac{\beta-1}{\tau}
-\left({{\beta}/{\pi}}\right)^{1/2}\left[\frac{\beta-2}{\sqrt{\tau}\tau}
-\frac{(\beta-1)(\beta-3)}{2\sqrt{\tau}\tau^3}
\right]\right\}.
\label{msdl2}
\end{eqnarray}
equation~(\ref{msdl2}) agrees with the standard diffusive.

On can see in figure~(\ref{fig8}) that there are a few particular points.

\begin{enumerate}
\item {$\beta=0$, ($M_\mathrm{s}=0$).} Substitution of this value of the parameter
$\beta$ into equation~(\ref{al}) yields $\alpha_{1,2}=\pm \ri$ and equation~(\ref{msd1})
is reduced to equation~(\ref{vartlg}).
\item { $\beta=3$ ($M=M_\mathrm{s}$)}. In this case, $\alpha_{1,2}=(\sqrt(3)\mp \ri)/2$
and the mean square displacement
is defined by the error function of complex argument.
\item {  $\beta=4$ ($M/M_\mathrm{s}=5/8$).} Substitution of $\beta=4$ into
equation~(\ref{al}) yields $\alpha_{1(2)}=1$.
It follows from equation~(\ref{msd1}) that
\begin{eqnarray}
\lim_{\alpha_1\to\alpha_2=1}\langle x^2(\tau)\rangle=
2D\tau_\mathrm{F}\left[\tau-6\sqrt{\frac{\tau}{\pi}}+3+(2\tau-3)
\re^\tau \mathrm{erfc}\left(\sqrt{\tau}\right)\right].
\label{msd3}
\end{eqnarray}
\item {  $\beta=9$ ($M=0$)}. In this case, the real parameters
$\alpha_{1,2}=(3\mp \sqrt{5})/2$.
\end{enumerate}

The mean square displacements for different values of the parameter $\beta$
listed above are shown in figure~\ref{fig9}. The hydrodynamic interaction
results in time delay of the mean square displacement. The larger is the value of
the parameter $\beta$, the slower is the convergence to the asymptotic limit.

It is necessary to note that the predictions of equation~(\ref{msd1}) are in a good
agreement with experimental data~\cite{LJTKFF05}

The Laplace transform of the mean square displacement given by equation~(\ref{msd1})
 reads
\begin{equation}
\langle \wt{\Delta \xi^2(s)}\rangle=
\frac{2}{s^2\left(s+1+\sqrt{\beta s}\right)}\,,
\label{msd1s}
\end{equation}
therefore, the memory kernel of the Langevin equation is given by
\begin{equation}
\wt{\gamma}(s)=1+\sqrt{\beta s}\,.
\label{ltkH}
\end{equation}
Substitution of equation~(\ref{ltkH}) to equation~(\ref{Lkernel}) yields the kernel
of the time-nonlocal Fokker-Planck equation with  due regard for
hydrodynamics effects
\begin{equation}
\wt{W}(s)=\frac{1}{s+1+\sqrt{\beta s}}\,.
\label{HkernFP}
\end{equation}

\subsection{Probability density function}
Solutions of the telegraph equation in higher than one dimension were
considered in~\cite{PMW97} and it was shown that the result is unphysical, in
some regions the PDF becomes negative. Nevertheless, a
reasonable PDF in explicit form was obtained for two-dimensional
systems within the framework of a persistent random walk model~\cite{MPW93}.
Therefore, the possibility to describe the high dimension diffusion process with
finite speed of propagation using the time-nonlocal approach is an open question.

In the general case, equation~(\ref{FPpdf}) reads in dimensionless units
\begin{equation}
\frac{\partial f(\xi,\tau)}{\partial \tau} = \int_0^\tau  W(\tau-\tau^\prime)
\Delta_d f(\xi,\tau^\prime)\ud \tau^\prime
\ ,
\label{dFPpdf}
\end{equation}
where $d$ is the space dimension and the Laplace operator is given by
\begin{equation}
\Delta_d=\frac{1}{\xi^{d-1}}\frac{\partial}{\partial \xi}\xi^{d-1}
\frac{\partial}{\partial \xi} =\frac{\partial^2}{\partial \xi^2}+\frac{d-1}{\xi}\frac{\partial}
{\partial \xi}\,.
\label{lapl}
\end{equation}
For pure ballistic propagation, the kernel in equation~(\ref{dFPpdf}) is defined
by $W(\tau)\equiv 1$. The time derivative of equation~(\ref{dFPpdf}) with this
kernel yields d-dimensional wave equation
\begin{equation}
\frac{\partial^2 f(\xi,\tau)}{\partial \tau^2}-\Delta_d f(\xi,\tau)=0.
\label{weq}
\end{equation}
It is clear that the propagation of the ballistic impulse is specified by the function
\begin{equation}
f(\xi,\tau)=\frac{\delta (\tau-\xi)}{\xi^{d-1}}\,,
\label{PM}
\end{equation}
where the normalization condition is defined by
$\int_0^{\infty} f(\xi,\tau)  \xi^{d-1} \ud \xi =1$.
Nevertheless, this function is the solution of equation~(\ref{weq}), i.e., the
one-dimensional case only; for higher dimensions, the solution differs from the
function defined by equation~(\ref{PM}), turning negative in some regions. This
is the source of the failure of the telegraph equation for $d>1$.
Substitution of equation~(\ref{PM}) into equation~(\ref{weq}) yields
\begin{equation}
\bigg(\frac{\partial^2 }{\partial \tau^2}
-\Delta_d\bigg) \frac{\delta (\tau-\xi)}{r^{d-1}}=
\frac{d-1}{\xi}\Bigg(\frac{\partial}{\partial \xi}
+\frac{d-2}{\xi}\Big)\frac{\delta (\tau-\xi)}{\xi^{d-1}}\,.
\label{pmwe}
\end{equation}
Therefore, we can suggest that the auxiliary equation which corresponds to
equation~(\ref{pmwe}) should be given by
\begin{eqnarray}
&&\frac{\partial f(\xi,\tau)}{\partial \tau}=
\frac{\partial^2 f(\xi,\tau)}{\partial \xi^2}+
2\frac{d-1}{\xi}\frac{\partial f(\xi,\tau)}{\partial \xi}+\frac{(d-1)(d-2)}{\xi^2}f(\xi,\tau).
\label{APMeq}
\end{eqnarray}

The Laplace transform of equation~(\ref{APMeq}) reads
\begin{equation}
\frac{\partial^2 \wt{f}(\xi,s)}{\partial \xi^2}+
2\frac{d-1}{\xi}\frac{\partial \wt{f}(\xi,s)}{\partial \xi}
+\left[\frac{(d-1)(d-2)}{\xi^2}-s\right]\wt{f}(\xi,s)=0.
\label{LAPMeq}
\end{equation}
This equation can be reduced to the Bessel equation~\cite{K4} and its
normalized solution is given by
\begin{equation}
\wt{f}(\xi,s)=\sqrt{\frac{2}{\pi}}\xi^{2-d}\frac{K_{1/2}\left(\xi\sqrt{s}\right)}
{\left(\xi\sqrt{s}\right)^{1/2}}\,.
\label{PDF1}
\end{equation}
The inverse Laplace transform of equation~(\ref{PDF1}) yields the PDF in
 the time domain
\begin{equation}
f(\xi,\tau)=\frac{1}{\sqrt{\pi}}\xi^{1-d}\frac{\re^{-\xi^2/(4\tau)}}{\sqrt{\tau}}\,.
\label{pdfnew}
\end{equation}
The PDF given by this equation differs from that of standard diffusion. Nevertheless,
the second moment of this function is in agreement with equation~(\ref{var1})
and the kurtosis of this distribution is given by $\kappa=3$.

We suggest that the time-nonlocal Fokker-Planck equation should be given by
\begin{equation}
\frac{\partial f(\xi,\tau)}{\partial \tau} = \int_0^\tau  W(\tau-\tau^\prime)
\hat{\mathcal{D}} f(\xi,\tau^\prime)\ud \tau^\prime,
\label{DtnFP}
\end{equation}
where the operator $\hat{\mathcal{D}}$ is defined by
\begin{equation}
\hat{\mathcal{D}}=\frac{\partial^2 }{\partial \xi^2}+
2\frac{d-1}{\xi}\frac{\partial }{\partial \xi}+
\frac{(d-1)(d-2)}{\xi^2} \ .
\label{Doper}
\end{equation}
The formal solution of equation~(\ref{DtnFP}) follows from equation~(\ref{PDF1})
and equation~(\ref{Lkernel})
\begin{equation}
\wt{f}(\xi,s)=\sqrt{\frac{2}{\pi}}\xi^{2-d}\left[s+\wt{\gamma}(s)\right]\frac{K_{1/2}
\left(\xi\sqrt{s\left[s+\wt{\gamma}(s)\right]}\right)}
{\left(\xi\sqrt{s\left[s+\wt{\gamma}(s)\right]}\right)^{1/2}}\,.
\label{FsDtnFP}
\end{equation}
This equation is similar to equation~(\ref{solLkern}) to within the coefficient
which depends on $\xi$.  Therefore, the nonvanishing part of the PDF for $d=3$
and the memory kernel defined by equation~(\ref{ltkH}) is given
[see equation~(\ref{extrsolv}) and equation~(\ref{f1s})] by
\begin{equation}
\wt{f}_{I}(\xi,s)=\frac{1}{\xi^2}
\exp\left[-r\left(s+\frac{1+\sqrt{\beta s}}{2}\right)\right].
\label{SpartH}
\end{equation}
For $\beta=0$, the inverse Laplace transform of this equation yields
the damped impulse propagation corresponding to equation~(\ref{PM}).  In a general
case, the inverse Laplace transform of equation~(\ref{SpartH}) reads
\begin{equation}
{f}_{I}(\xi,\tau)=\frac{1}{4\xi}\re^{-\xi/2}\sqrt{\frac{\beta}{\pi}}
\frac{\exp\left\{-\beta\xi^2/\left[16(\tau-\xi)\right]\right\}}{\sqrt{(\tau-\xi)^3}}\,.
\label{SpartTH}
\end{equation}

\begin{figure}[!t]
\centering
\includegraphics[width=0.4\textwidth]{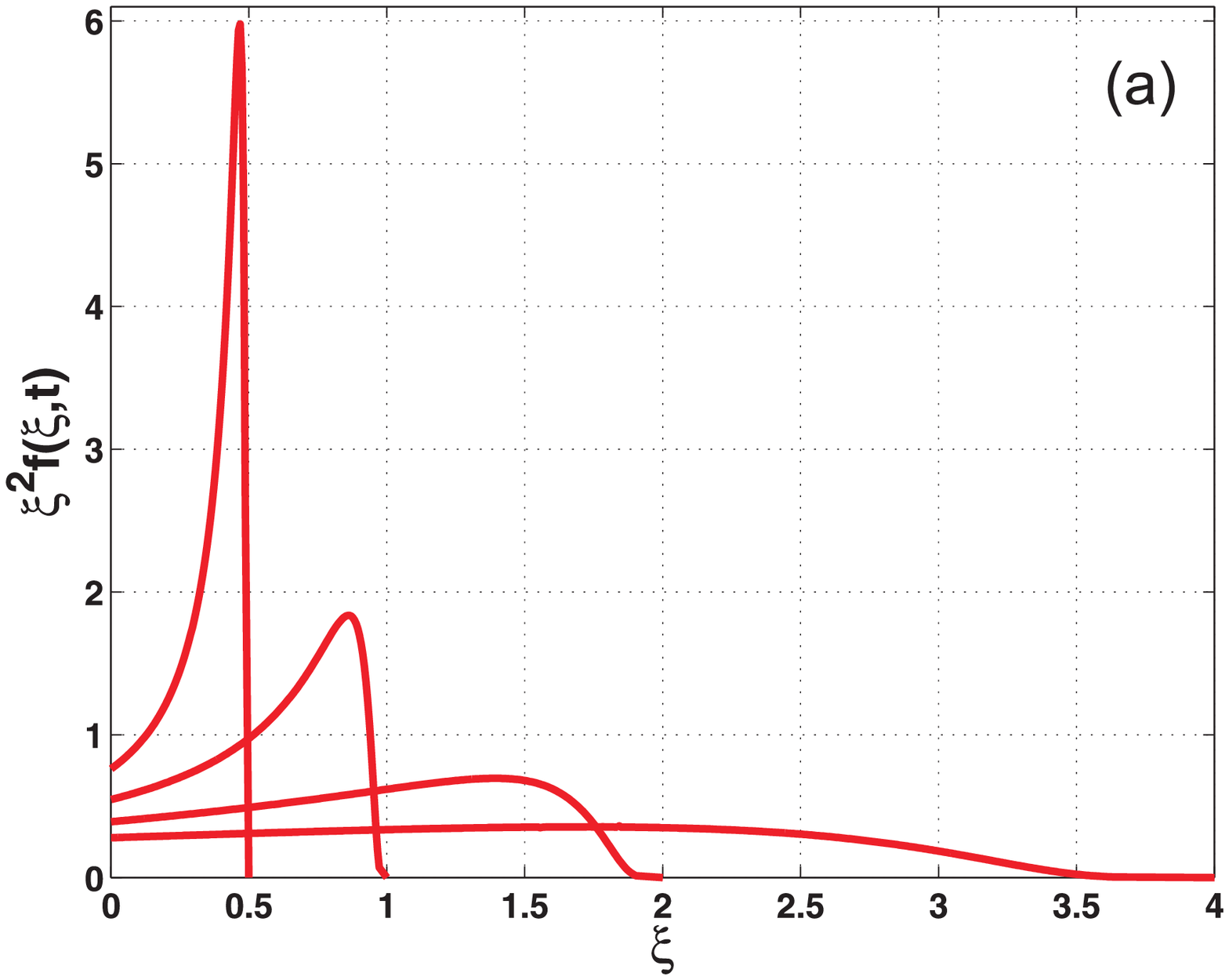}
\hspace{1cm}
\includegraphics[width=0.4\textwidth]{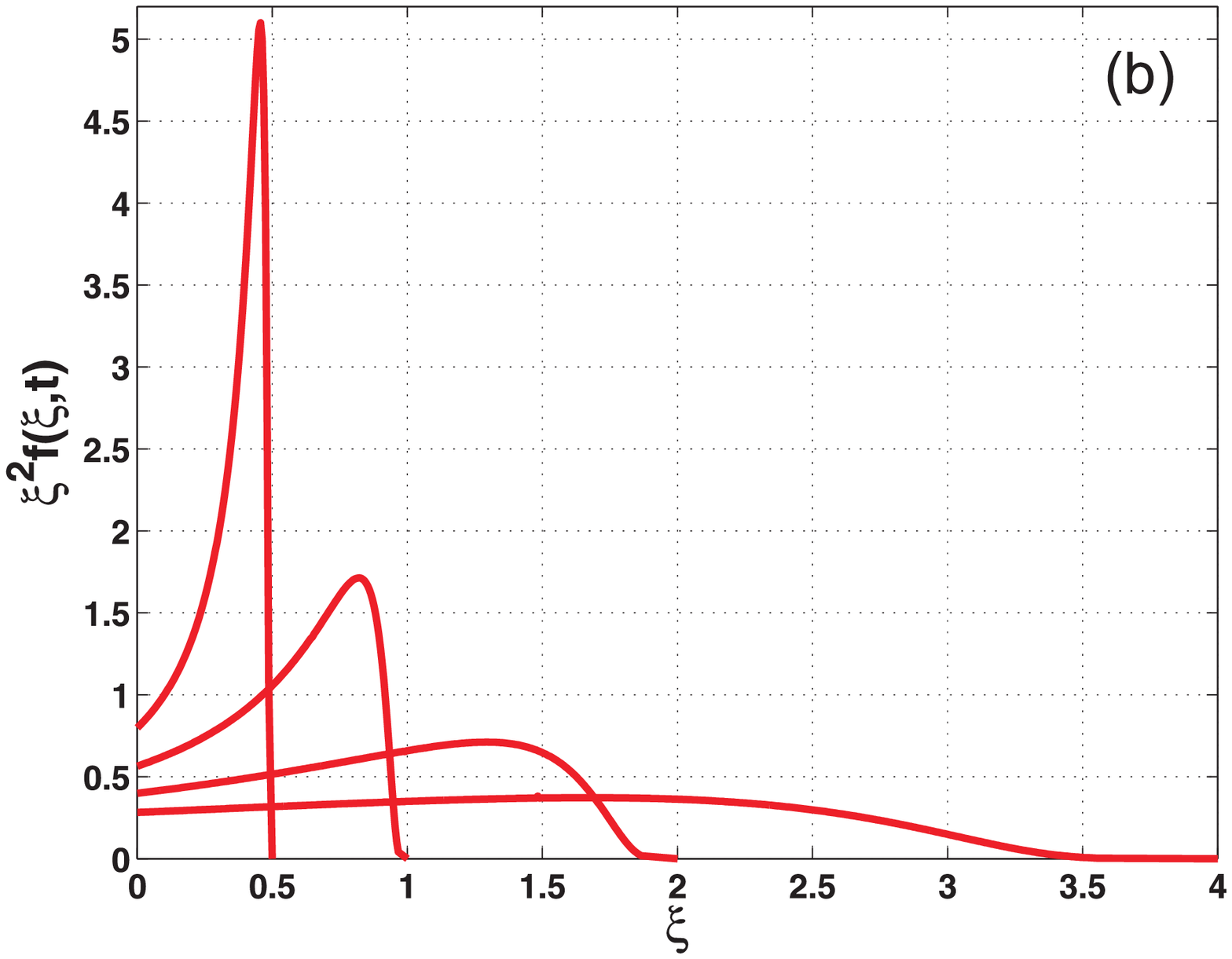}\\
\includegraphics[width=0.4\textwidth]{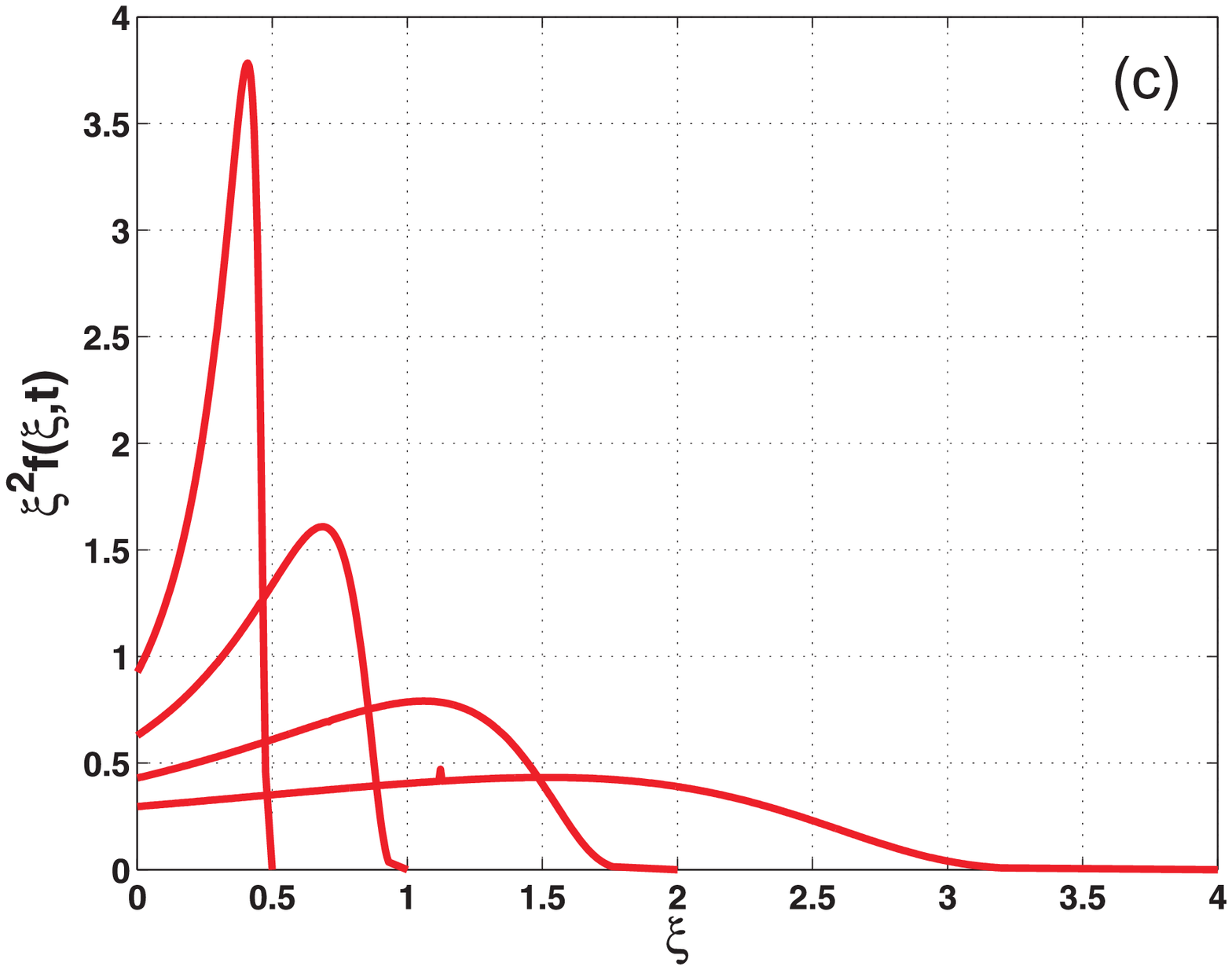}
\caption{(Color online) The PDF defined by equation~(\ref{HfullTsol}) for $t=0.5,1,2,4$.
Panel (a): $\beta=3$; Panel (b): $\beta=4$; Panel (c): $\beta=9$
}
\label{fig10}
\end{figure}

\begin{figure}[!t]
\centering
\includegraphics[width=0.4\textwidth]{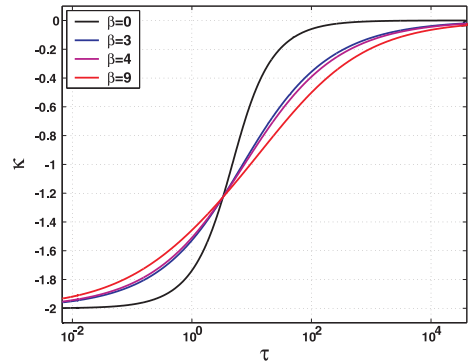}
\caption{(Color online) The time dependence of the kurtosis of the PDF corresponding
to the memory kernel defined by equation~(\ref{HkernFP}).}
\label{fig11}
\end{figure}

In the limit $\tau \to 0$, the function defined by equation~(\ref{SpartTH}) tends to
three-dimensional delta-function, i.e., the initial condition of the problem under
consideration. In infinitesimal time, the delta-function due to hydrodynamic
interactions becomes smooth and the PDF is free from the singular
contribution. Therefore, there is no need to pull out the ballistic
contribution from the full PDF, and the solution of the problem is given by
\begin{equation}
f(\xi,\tau)=\Phi(\xi,\tau-\xi)\Theta(\tau-\xi),
\label{HfullTsol}
\end{equation}
where the inverse Laplace transform of the continuous function is
\begin{eqnarray}
\wt{\Phi}(\xi,s)&=&\left[{1+{1}/{s}+\left({{\beta}/{s}}\right)^{1/2}}\right]^{1/2}
\exp\left(-\mid \xi\mid\cdot s\left\{\left[{1+{1}/{s}+\left({{\beta}/{s}}\right)^{1/2}}\right]^{1/2}-1\right\}\right)\,.
\label{pdfspr}
\end{eqnarray}
This can be estimated numerically. The results of calculations for different
values of the parameter $\beta$ are shown in figure~\ref{fig10}.
Note that a similar ballistic peak of the PDF was observed
in simulations of the test particle transported in pseudoturbulent fields~\cite{HJE07}.
The kurtosis of the PDF for different values of the parameter $\beta$ are
compared in figure~\ref{fig11}.

\section{Conclusion}
The strategy followed in this paper is to construct a time-nonlocal Fokker-Planck
equation which reproduces the time dependence of the mean square displacement of
an underlying process throughout the time domain. It should be stressed that
the same mean square displacement may correspond, in
general, to different models for the time-dependence of the PDF. Thus, the predictions of the time non-local Fokker-Planck
approach should be compared to experimental data for the PDF to assess its scope of validity and
 the quality of the approximation. The advantage of the proposed model is that it can deal with diffusion processes that cross-over
from a ballistic to a fractional behavior when time increases from short to long
values, respectively.

The general one-dimensional solution
(\ref{fullsolb}) demonstrates the effect of the temporal memory in the form of
a partition of the probability distribution function inside a growing spatial domain
which increases in a deterministic way. The approach provides a solution that
exists at all times, and, in particular, is free from  the instantaneous
action puzzle. An extension of the employed approach to higher spatial
dimensions is used to study the implications of hydrodynamic interactions on
the shape of the PDF. It is shown that singular ballistic contribution to the PDF is smoothed out during the propagation. The expansion given by
equation~(\ref{solPsiser}) splits the Laplace transform of the PDF into two parts.
One part is the ballistic part which is the solution of an inhomogeneous differential
equation in spatial variable with the initial condition of the problem. The second one is the
diffusion part, which solved the homogeneous equation and
is zero at the initial time. In its turn, the diffusion part is divided again into a few
terms which correspond, in general, to different auxiliary equations.
Therefore, each diffusion term in $s$-space can be  multiplied by a
coefficient $C(s)$ so that the normalization condition and the mean square
displacement are preserved. Moreover, there is a freedom of modelling here, and this freedom has not been
exhausted in this paper. Further analysis and comparison with experimental systems are necessary
to nail down the various options of modeling different physical realizations. This research will
be the subject of the coming publications.

\ukrainianpart

\title{Рівняння Фокера-Планка з пам'яттю: кросовер від балістичного до дифузійного процесу
в багаточастинкових системах і нестисних середовищах}

\author{В. Ільїн\refaddr{ad1}, І. Прокаччіа\refaddr{ad1}, A. Загородній\refaddr{ad2}}
\addresses{ \addr{ad1} Відділ хімічної фізики, Науковий Інститут Вайцмана, 76100 Реховот, Ізраїль
\addr{ad2} Інститут теоретичної фізики ім. М.М. Боголюбова НАН України, 03680 Київ, Україна
}

\makeukrtitle

\begin{abstract}
На основі немарківського узагальнення рівняння Фокера-Планка запропоновано підхід до об'єднаного опису дифузійних процесів, який дозволяє розглядати як балістичний режим на малих часах, так і аномальну (суб- або супер-) дифузію на великих часових інтервалах. Встановлено зв'язок немарківських кінетичних коефіцієнтів зі спостережуваними величинами (середіми та середньоквадратичними зміщеннями). Отримано розв'язки, що описують дифузійні процеси у фізичному просторі. Для великих часів еволюції вони узгоджуються з результатами теорії неперервних в часі випадкових блукань, а на малих часах описують балістичну динаміку.

\keywords немарківські процеси, дробова дифузія, балістичні ефекти
\end{abstract}


\begin{thebibliography}{99}

\bibitem{P05} Philibert J.,  Diffusion Fundamentals, 2005, \textbf{2}, 1.

\bibitem{51Hur}
 Hurst H.E.,  Trans. Am. Soc. Civil Eng., 1951, \textbf{116}, 770.


\bibitem{MN68}  Mandelbrot B.B.,  van Ness J.W.,  SIAM Rev., 1968, \textbf{10}, 422; \doi{10.1137/1010093}.


\bibitem{Ein1905} {Einstein A.,  Ann. Phys., 1905, \textbf{17}, 549.}


\bibitem{Ein1906} {Einstein A., Ann. Phys., 1906, \textbf{19}, 371.}


\bibitem{L908} Langevin P., C. R. Acad. Sci. (Paris), 1908, \textbf{146}, 530.

\bibitem{LG97} {Lemons D.S.,  Gythiel A., Am. J. Phys., 1997, \textbf{65}, 1079; \doi{10.1119/1.18725}.}

\bibitem{UO30} Uhlenback G.E.,  Ornstein L.S.,  Phys. Rev., 1930, \textbf{36}, 823¦ \doi{10.1103/PhysRev.36.823}.

\bibitem{LJTKFF05} Luki\'c B.,  Jeney S.,  Tischer C.,  Kulik A.J.,  Forr\'o L.,
 Florin E.L.,  Phys. Rev. Lett., 2005, \textbf{95}, 160601; \\ \doi{10.1103/PhysRevLett.95.160601}.

\bibitem{HCTLJRF11}  Huang R.,  Chavez I.,  Taute K.M.,   Luki\'c B.,  Jeney S.,
 Raizen M.G.,  Florin E.L.,  Nat. Phys., 2011, \textbf{7},
  576; \\ \doi{10.1038/nphys1953}.

\bibitem{P11}  Pusey P.N.,  Science, 2011, \textbf{332}, 802; \doi{10.1126/science.1192222}.

\bibitem{D34} Davydov B.I.,  Dokl. Akad. Nauk SSSR, 1934, \textbf{2},  474.

\bibitem{MF53} Morse P.M.,  Feshbach H., {Methods of Theoretical
Physics}, McGraw-Hill, New York, 1953.

\bibitem{IPZ10} Ilyin V.,  Procaccia I.,  Zagorodny A.,
Phys. Rev. E, 2010, \textbf{81}, 030105(R); \doi{10.1103/PhysRevE.81.030105}.

\bibitem{IZ10} Ilyin V.,   Zagorodny A.,  Ukr. J. Phys., 2010,
\textbf{55}, 235.


\bibitem{VT45} Vladimirsky V.,  Terletzky Y.A., Zh. Eksp. Teor. Fiz., 1945, \textbf{15}, 258 (in Russian).


 \bibitem{C48} Cattaneo C., Atti. Sem. Mat. Fis. Univ. Modena, 1948, \textbf{3}, 83.

\bibitem{APLA08} Reverberia A.P., Bagnerini P., Maga L., Bruzzone A.G.,   Int. J. Heat Mass Transfer, 2008, \textbf{51}, 5327; \\ \doi{10.1016/j.ijheatmasstransfer.2008.01.039}.

\bibitem{S02} Sokolov I.M.,  Phys. Rev. E, 2002, \textbf{66}, 041101; \doi{10.1103/PhysRevE.66.041101}.

\bibitem{M65} H. Mori, Prog. Theor. Phys., 1965, \textbf{33}, 423; \doi{10.1143/PTP.33.423}.

\bibitem{KI66}R. Kubo, The Fluctuation-Dissipation Theorem and Brownian
Motion, 1965 Tokyo Summer Lectures in Theoretical Physics,
Part 1. Many-Body Theory, Kubo R. (Ed.), Syokabo, Tokyo and
Benjamin, New York, 1966, 1--16.

\bibitem{K66} Kubo R.,  Rep. Prog. Phys., 1966, \textbf{29}, 255; \doi{10.1088/0034-4885/29/1/306}.

\bibitem{L05} Luczka J., Chaos, 2005, \textbf{15}, 026107; \doi{10.1063/1.1860471}.

\bibitem{PWM96} Porr\`a J.M., Wang K.-G.,  Masoliver J.,
 Phys. Rev. E, 1996, \textbf{53}, 5872; \doi{10.1103/PhysRevE.53.5872}.

\bibitem{duf93} Duffy D.G., ACM TOMS, 1993, \textbf{19}, 333; \doi{10.1145/155743.155788}.

\bibitem{MK00} Metzler R.,  Klafter J.,  Phys. Rep., 2000, \textbf{339}, 1; \doi{10.1016/S0370-1573(00)00070-3}.

\bibitem{BHW87}  Ball R.C.,  Havlin S.,  Weiss G.H.,
J. Phys. A: Math. Gen., 1987, \textbf{20}, 4055; \doi{10.1088/0305-4470/20/12/052}.

\bibitem{AS}  {Handbook of Mathematical Functions with Formulas, Graphs, and Mathematical Tables},  Abramowitz M.,  Stegun I.A. (Eds.), Dover, New York, 1972.


\bibitem{B885} Boussinesq J.V.,  C. R. Acad. Sci., 1885, \textbf{100}, 935.


\bibitem{B61} Basset A.B., Treatise on Hydrodynamics., Vol. 2,
Dover, New York, 1961, Chap.22.


\bibitem{LL87} Landau L.D.,  Lifshitz E.M., {Fluid Mechanics},
Pergamon Press, Oxford, New York, 1987.

\bibitem{MP96} Mainardi F.,  Pironi P.,
Extr. Math., 1996, \textbf{11}, 140.

\bibitem{PMW97} Porr\`a J.M.,  Masoliver J.,  Weiss G.H., Phys. Rev. E, 1997, \textbf{55}, 7771; \doi{10.1103/PhysRevE.55.7771}.

\bibitem{MPW93} Masoliver J.,  Porr\`a J.M.,   Weiss G.H.,  Physica A,  1993, \textbf{193}, 469; \doi{10.1016/0378-4371(93)90488-P}.

\bibitem{K4} Kamke E., {Differentialgleichungen. L\"osungsmethoden und
L\"osungen I. Gew\"ohnliche Differentialgleichungen}, Leipzig, 1942.

\bibitem{HJE07} Hauff T.,  Jenko F.,  Eule S.,  Phys. Plasmas, 2007, \textbf{14}, 102316; \doi{10.1063/1.2794322}.

\end{thebibliography}
\end{document}